\documentclass[letterpaper,twocolumn,showpacs,
prb
,superscriptaddress,email
,showkeys%
]{revtex4}

\usepackage{soul}
\usepackage[pdftex]{graphicx}
\usepackage{amssymb}
\usepackage{mathrsfs}
\usepackage{amsmath,amsfonts,latexsym}
\usepackage{array,tabularx,color}
\usepackage{dcolumn} 
\usepackage[normalem]{ulem}
\usepackage{comment}
\usepackage{bm}

\begin{document}

\title{Energy relaxation of exciton-polariton condensates in quasi-1D microcavities}

\author{C. Ant\'on}
\affiliation{Departamento de F\'isica de Materiales, Universidad Aut\'onoma de Madrid, Madrid 28049, Spain}
\affiliation{Instituto de Ciencia de Materiales ``Nicol\'as Cabrera'', Universidad Aut\'onoma de Madrid, Madrid 28049, Spain}

\author{T. C .H. Liew}
\affiliation{School of Physical and Mathematical Sciences, Nanyang Technological University, 637371, Singapore}

\author{G. Tosi}
\affiliation{Departamento de F\'isica de Materiales, Universidad Aut\'onoma de Madrid, Madrid 28049, Spain}

\author{M. D. Mart\'in}
\affiliation{Departamento de F\'isica de Materiales, Universidad Aut\'onoma de Madrid, Madrid 28049, Spain}
\affiliation{Instituto de Ciencia de Materiales ``Nicol\'as Cabrera'', Universidad Aut\'onoma de Madrid, Madrid 28049, Spain}

\author{T. Gao}
\affiliation{Department of Materials Science and Technology, Univ. of Crete, 71003 Heraklion, Crete, Greece}
\affiliation{FORTH-IESL, P.O. Box 1385, 71110 Heraklion, Crete, Greece}

\author{Z. Hatzopoulos}
\affiliation{FORTH-IESL, P.O. Box 1385, 71110 Heraklion, Crete, Greece}
\affiliation{Department of Physics, University of Crete, 71003 Heraklion, Crete, Greece}

\author{P. S. Eldridge}
\affiliation{FORTH-IESL, P.O. Box 1385, 71110 Heraklion, Crete, Greece}

\author{P. G. Savvidis}
\affiliation{Department of Materials Science and Technology, Univ. of Crete, 71003 Heraklion, Crete, Greece}
\affiliation{FORTH-IESL, P.O. Box 1385, 71110 Heraklion, Crete, Greece}

\author{L. Vi\~na}
\email{luis.vina@uam.es}
\affiliation{Departamento de F\'isica de Materiales, Universidad Aut\'onoma de Madrid, Madrid 28049, Spain}
\affiliation{Instituto de Ciencia de Materiales ``Nicol\'as Cabrera'', Universidad Aut\'onoma de Madrid, Madrid 28049, Spain}
\affiliation{Instituto de F\'isica de la Materia Condensada, Universidad Aut\'onoma de Madrid, Madrid 28049, Spain}


\begin{abstract}
We present a time-resolved study of energy relaxation and trapping dynamics of polariton condensates in a semiconductor microcavity ridge. The combination of two non-resonant, pulsed laser sources in a GaAs ridge-shaped microcavity gives rise to profuse quantum phenomena where the repulsive potentials created by the lasers allow the modulation and control of the polariton flow. We analyze in detail the dependence of the dynamics on the power of both lasers and determine the optimum conditions for realizing an all-optical polariton condensate transistor switch. The experimental results are interpreted in the light of simulations based on a generalized Gross-Pitaevskii equation, including incoherent pumping, decay and energy relaxation within the condensate.
\end{abstract}



\maketitle

\section{Introduction}
\label{sec:intro}

Bose-Einstein condensation of quasiparticles in solid-state systems has been observed in excitons in quantum Hall bilayers,~\cite{Eisenstein2004} exciton-polaritons in semiconductor microcavities,~\cite{Kasprzak:2006fk} gases of magnons,~\cite{Demokritov:2006qy,Giamarchi:2008uq} cavity photons~\cite{Klaers:2010kx} and indirect excitons.~\cite{High:2012fk} Exciton-polaritons, mixed light-matter quasiparticles behaving as bosons, form condensates which exhibit not only the fundamental properties of quantum gases, but also new fascinating phenomena related to their out-of-equilibrium character.~\cite{Carusotto:2013uq} The photonic component of polaritons is responsible for their light mass, which makes condensation possible up to room temperature~\cite{chris2007}, and for their easy creation, manipulation and detection by using simple optical-microscopy setups. On the other hand, their excitonic component yields strong Coulomb repulsive interactions that make them promising candidates for future non-linear optical technologies.

The peculiar quantum fluid properties of polariton condensates are under intense research nowadays. Recent findings include: robust propagation of coherent polariton bullets~\cite{Amo2009} and elucidation of the validity of the Landau criterion for frictionless flow in the presence of weak structural defects,~\cite{Amo2009a} persistent quantized superfluid rotation,~\cite{Sanvitto2010,Marchetti2010} and solitary waves resulting from compensation between dispersion and particle interaction.~\cite{Amo:2011qf,SichM.:2012ul,Grosso2011} Moreover, the intrinsic out-of-equilibrium character of polariton condensates has motivated recent theoretical studies on how to describe properly the energy flow from an optically-injected hot exciton reservoir to the coherent polariton modes,~\cite{RevModPhys.82.1489,1367-2630-14-7-075020} which we carefully address in this work.

The functionalities of microcavities in the strong coupling regime, as integrated optical elements, promote polaritons as an undreamt platform to create new logical devices.~\cite{Liew2011} Thanks to their interactions with non-condensed excitons, polaritons can be easily accelerated, propagating over macroscopic distances in high finesse microcavities.~\cite{Wertz2010,Christmann:2012hc} In this case, new interferometric devices can be built by properly shaping the excitation profile~\cite{Tosi2012} as well as the microcavity etching.~\cite{Sturm2013}

Extra confinement can be achieved by lateral bounding the optical modes through patterning the microcavity,~\cite{Galbiati:2012bh} by sculpting the pumping profile creating blueshift-induced traps,~\cite{Amo:2010kl,Tosi:2012zr} or by a combination of both methods.~\cite{Wertz2010,Wertz:2012ee,Tsotsis:2012qy,Gao:2012tg,anton:261116} This paves the way for studies of atom-like scenarios in which the energy spectrum becomes discrete. In a recent work using quasi 1D-microwire ridges, a polariton condensate transistor switch has been realized through optical excitation with two beams.~\cite{Gao:2012tg,anton:261116} One of the beams creates a polariton condensate which serves as a source (\emph{S}) of polaritons; their propagation is gated using a second weaker gate beam (\emph{G}) that controls the polariton flow by creating a local blueshifted barrier (a list of symbols used in the manuscript are given in the Appendix A). The ON state of the transistor (absence of \emph{G}) corresponds to forming a trapped condensate at the edge of the ridge (collector, \emph{C}) labelled as $\mathscr{C}_{C}$. The presence of \emph{G} hinders the propagation of polaritons towards \emph{C}, remaining blocked between \emph{S} and \emph{G} (OFF state). An insight of the energy relaxation and dynamics of the condensed polariton propagation in this system has been obtained lately by a time-resolved study of the ON/OFF states.~\cite{anton:261116} In the present work, we make a systematic study of the influence of the density of polaritons created in \emph{S} and \emph{G}on the propagation and the gating of polariton bullets, of their energy and density relaxation and of the optimal conditions for realizing an all-optical polariton condensate transistor switch. Our experiments are compared with simulations of the polariton condensate dynamics based on a generalized Gross-Pitaevskii equation, modified to account for incoherent pumping, decay and energy relaxation within the condensate.

\section{Sample and experimental setup}
\label{sec:sample_setup}

\begin{figure}
\setlength{\abovecaptionskip}{-5pt}
\setlength{\belowcaptionskip}{-2pt}
\begin{center}
\includegraphics[trim=0.5cm 0.4cm 0.4cm 0.1cm, clip=true,width=1.0\linewidth,angle=0]{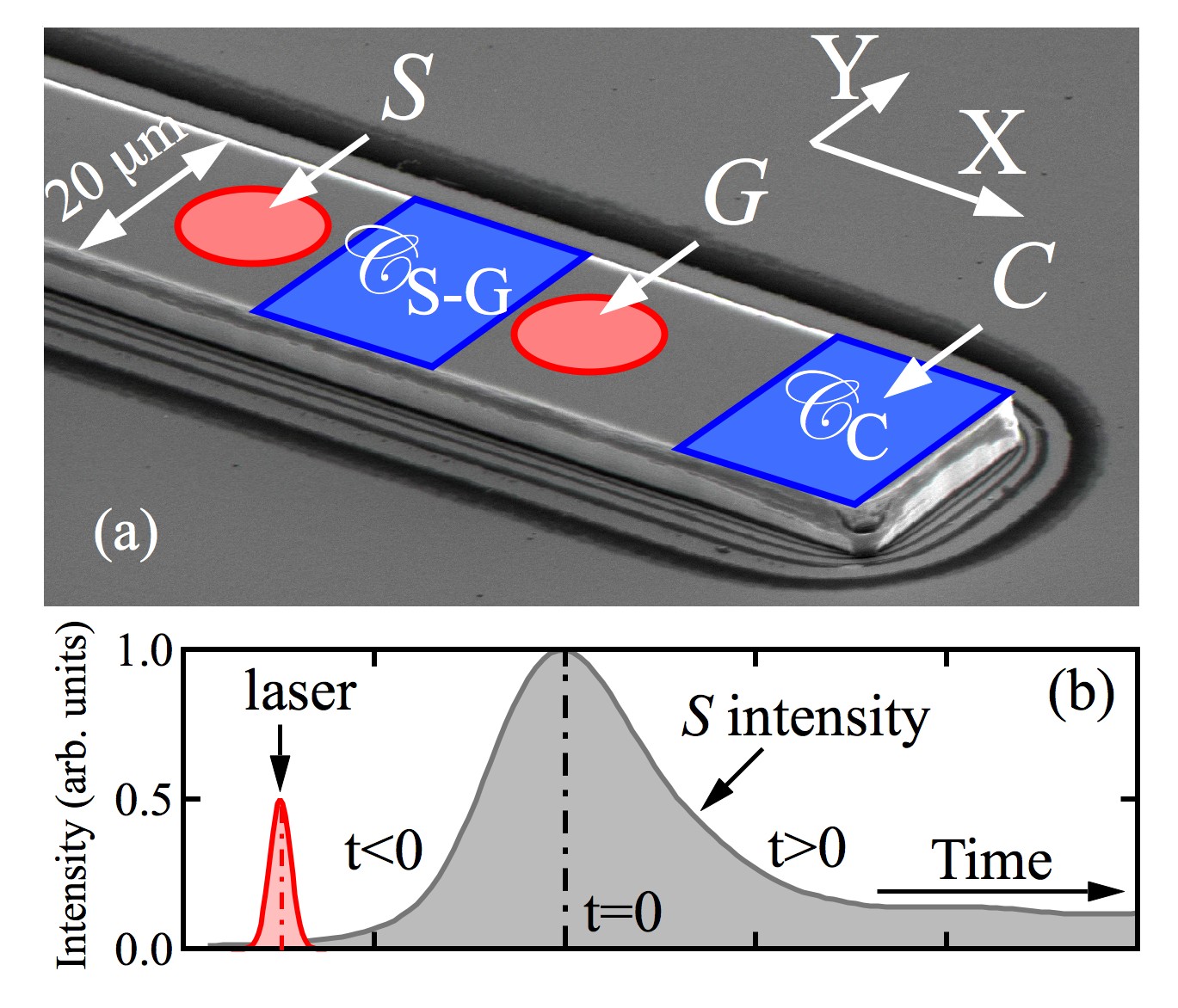}
\end{center}
\caption{(Color online) (a) Scanning electron microscopy image of a 20-$\mu$m wide ridge, including the excitation scheme with the \emph{S} and \emph{G} beams, and the position of the trapped condensates: $\mathscr{C}_{S-G}$, between \emph{S} and \emph{G}, and $\mathscr{C}_{C}$, at \emph{C}. (b) Temporal scheme to clarify the choice of the time origin: the instant $t=0$ is set at the maximum \emph{S}-intensity, the arrival of a given laser beam takes place at an instant $t<0$.}
\label{fig:fig0}
\end{figure}

\begin{figure}
\setlength{\abovecaptionskip}{-5pt}
\setlength{\belowcaptionskip}{-2pt}
\begin{center}
\includegraphics[trim=1.5cm 0.70cm 1.4cm 0.4cm, clip=true,width=1.0\linewidth,angle=0]{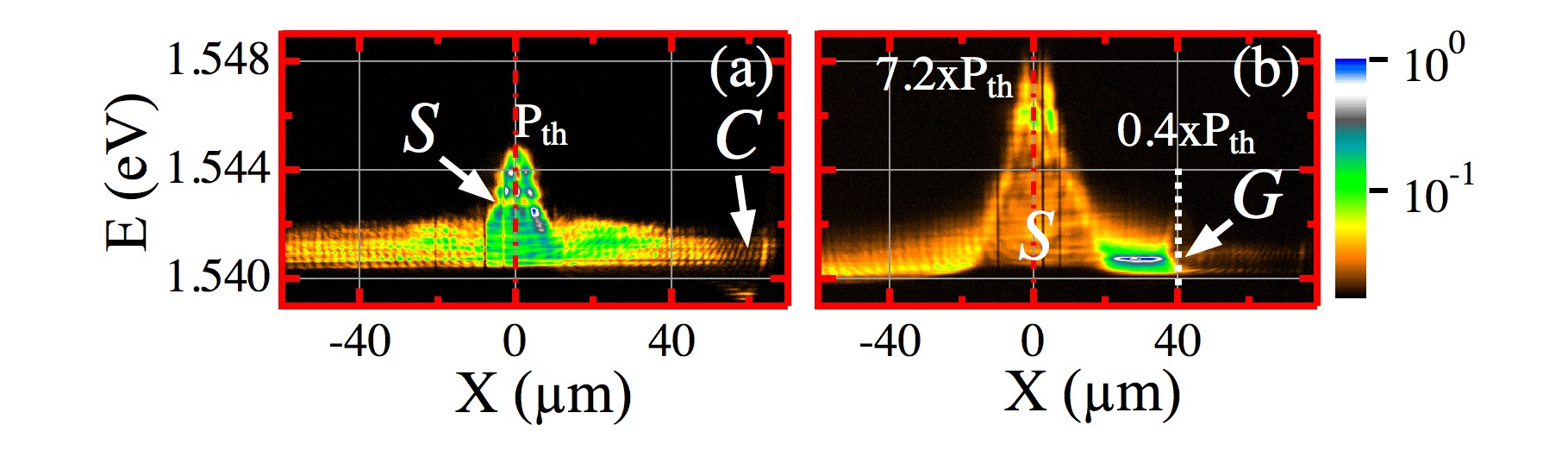}
\end{center}
\caption{(Color online) Energy vs. real space ($X$) image of a cross-section along the ridge under non-resonant CW-excitation. (a) Only one beam (\emph{S}) at $\backsim$75~$\mu$m from the right ridge border. (b) Two beam (\emph{S+G}) placed at $\backsim$75~$\mu$m and $\backsim$35~$\mu$m from the border, respectively; the dashed vertical line indicates the gate position. The intensity is coded in a logarithmic false color scale shown on the right.}
\label{fig:fig1}
\end{figure}

We investigate a high-quality $5\lambda/2$ AlGaAs-based microcavity with 12 embedded quantum wells, with a Rabi splitting of $\Omega_R = 9~$ meV. Ridges have been sculpted through reactive ion etching with dimensions $20 \times 300~\mu$m$^2$ (further information about this sample is given in Refs. \onlinecite{Tosi:2012zr} and \onlinecite{Tsotsis:2012qy}). Figure~\ref{fig:fig0} (a) shows a scanning electron microscopy image of such a ridge, including the excitation scheme; a temporal scheme of the excitation and emission processes is given in panel (b). 
In our sample lateral confinement is insignificant as compared to much thinner, 1D polariton wires.~\cite{Wertz2010,Wertz:2012ee} The chosen ridge is in a region of the sample corresponding to resonance (detuning between the bare exciton and bare cavity mode is $\delta \backsim$ 0). The sample, mounted in a cold-finger cryostat and kept at 10 K, is excited with 2 ps-long light pulses from a Ti:Al$_2$O$_3$ laser, tuned to the first high-energy Bragg mode of the microcavity (1.612 eV). We split the laser beam into two independent beams, whose intensities, spatial positions and relative time delay (zero for these experiments) can be independently adjusted. We focus both beams on the sample through a microscope objective to form 5~$\mu$m-$\varnothing$ spots spatially separated by $\backsim$40~$\mu$m along the ridge. The same objective is used to collect (angular range $\pm18^\circ$) and direct the emission towards a spectrometer coupled to a streak camera obtaining energy-, time- and spatial-resolved images, with resolutions of 0.4 meV, 15 ps and 1~$\mu$m, respectively. In our experiments polaritons propagate along the $X$ axis of the ridge. There is also some diffusion of polaritons in the $Y$ direction, but it is not relevant for the operation of our device. All the images in the manuscript show the emission collected along the $X$ axis from a 10-$\mu$m wide, central region of the ridge. The power threshold for condensation of polaritons is $P_{th}=1.5$ mW. Figure~\ref{fig:fig1} shows, as an example, under CW conditions, the intensity distribution of the polariton emission as a function of energy and of the position in the ridge: when we only use the \emph{S} beam, Fig.~\ref{fig:fig1} (a), we place it $\backsim$75~$\mu$m away from the right ridge border; in the \emph{S}+\emph{G} beam excitation, \emph{G} is placed $\backsim$35~$\mu$m away from the border, Fig.~\ref{fig:fig1} (b). In Fig.~\ref{fig:fig1} (a), exciting with a power $P_S=P_{th}$, the blue-shifted ($\backsim 3$ meV) emission at the source, together with a weak condensate emission $\mathscr{C}_{C}$, are clearly observed. $\mathscr{C}_{C}$ emits from an energy lower than that of the propagating polaritons as a result of an unintentional modification of the microcavity structure created by the etching process at the edge of the ridge. Polaritons propagate at a constant energy towards the left and right sides of the ridge. Exciting with both laser beams, under different excitation conditions for a typical OFF state, $P_S=7.2\times P_{th}$ and $P_G=0.4\times P_{th}$, the gating state of the switch is readily seen with the stopped, condensed polaritons just before the \emph{G} position. 

\section{Experimental results and discussion}
\label{sec:exp}

In this work, we time-resolve the different excitation configurations presented in Ref. \onlinecite{Gao:2012tg}: Fig. 2 (where the $P_S$ is varied whilst $P_G=0$) and Fig. 3 ($P_S=const.$ and $P_G$ is varied). Our study obtains intensity- and energy-dynamics of exciton and polariton emission in the ridge.
In the former case varying $P_S$, we fully characterize the ON state response of a polariton transistor switch; in the latter one, we modulate the polariton condensate trapping potential.~\cite{Wertz2010}

\subsection{One-beam excitation}
\label{subsec:one_beam}

\begin{figure}
\setlength{\abovecaptionskip}{-5pt}
\setlength{\belowcaptionskip}{-2pt}
\begin{center}
\includegraphics[trim=1.2cm 0.60cm 1.0cm 0.4cm, clip=true,width=1.0\linewidth,angle=0]{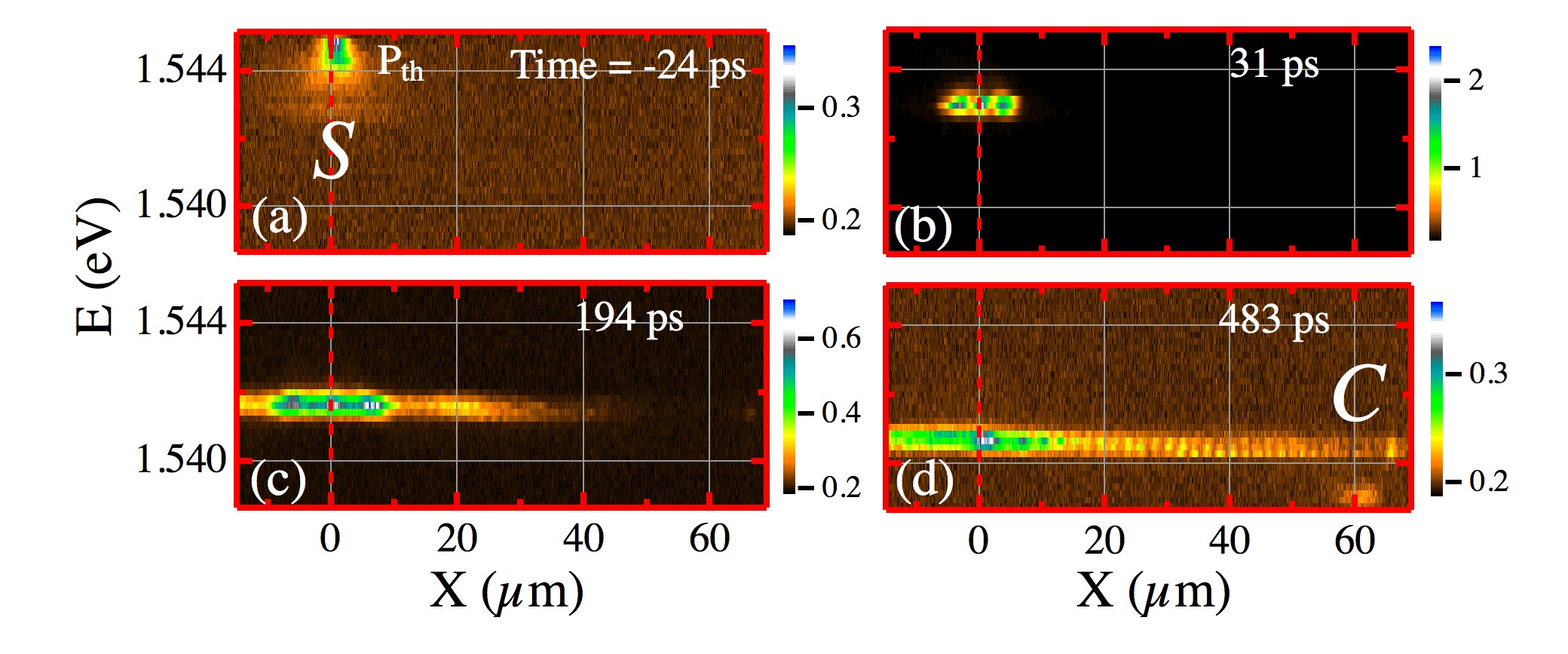}
\end{center}
\caption{(Color online) Energy vs. real space ($X$) for $P_S=P_{th}$ at different times shown by the labels. \emph{S} and \emph{C} mark the source and collector positions, respectively. The intensity is coded in a false color scale shown on the right of each panel. }
\label{fig:fig2}
\end{figure}

\begin{figure}
\setlength{\abovecaptionskip}{-5pt}
\setlength{\belowcaptionskip}{-2pt}
\begin{center}
\includegraphics[trim=1.2cm 0.6cm 0.7cm 0.4cm, clip=true,width=1.0\linewidth,angle=0]{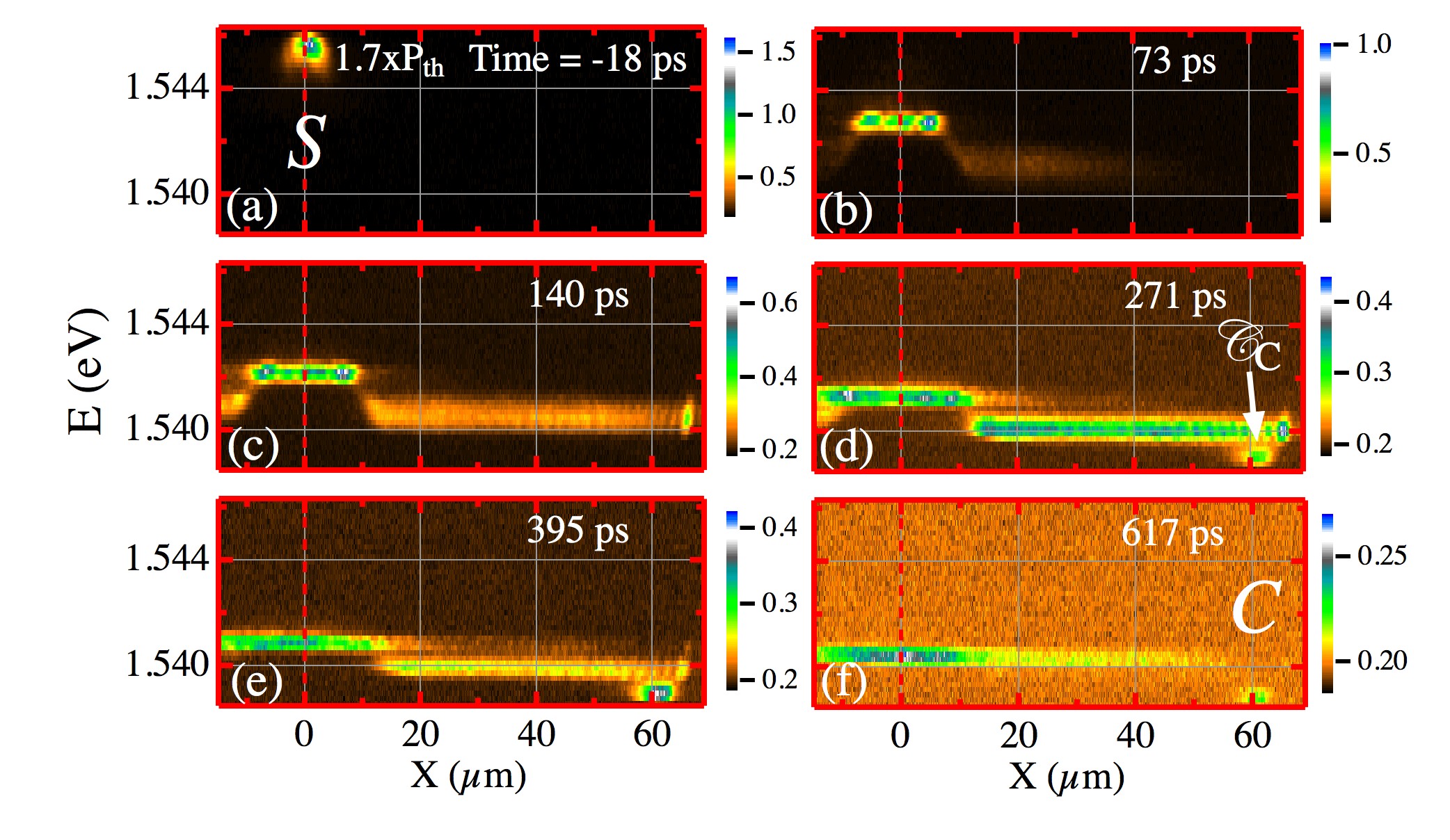}
\end{center}
\caption{(Color online) Energy vs. real space ($X$) for $P_S=1.7 \times P_{th}$ at different times shown by the labels. \emph{S}, \emph{C} and $\mathscr{C}_{C}$ mark the source, the collector and the trapped condensate positions, respectively. The intensity is coded in a false color scale shown on the right of each panel.}
\label{fig:fig3}
\end{figure}

\begin{figure}
\setlength{\abovecaptionskip}{-5pt}
\setlength{\belowcaptionskip}{-2pt}
\begin{center}
\includegraphics[trim=1.2cm 0.6cm 0.7cm 0.4cm, clip=true,width=1.0\linewidth,angle=0]{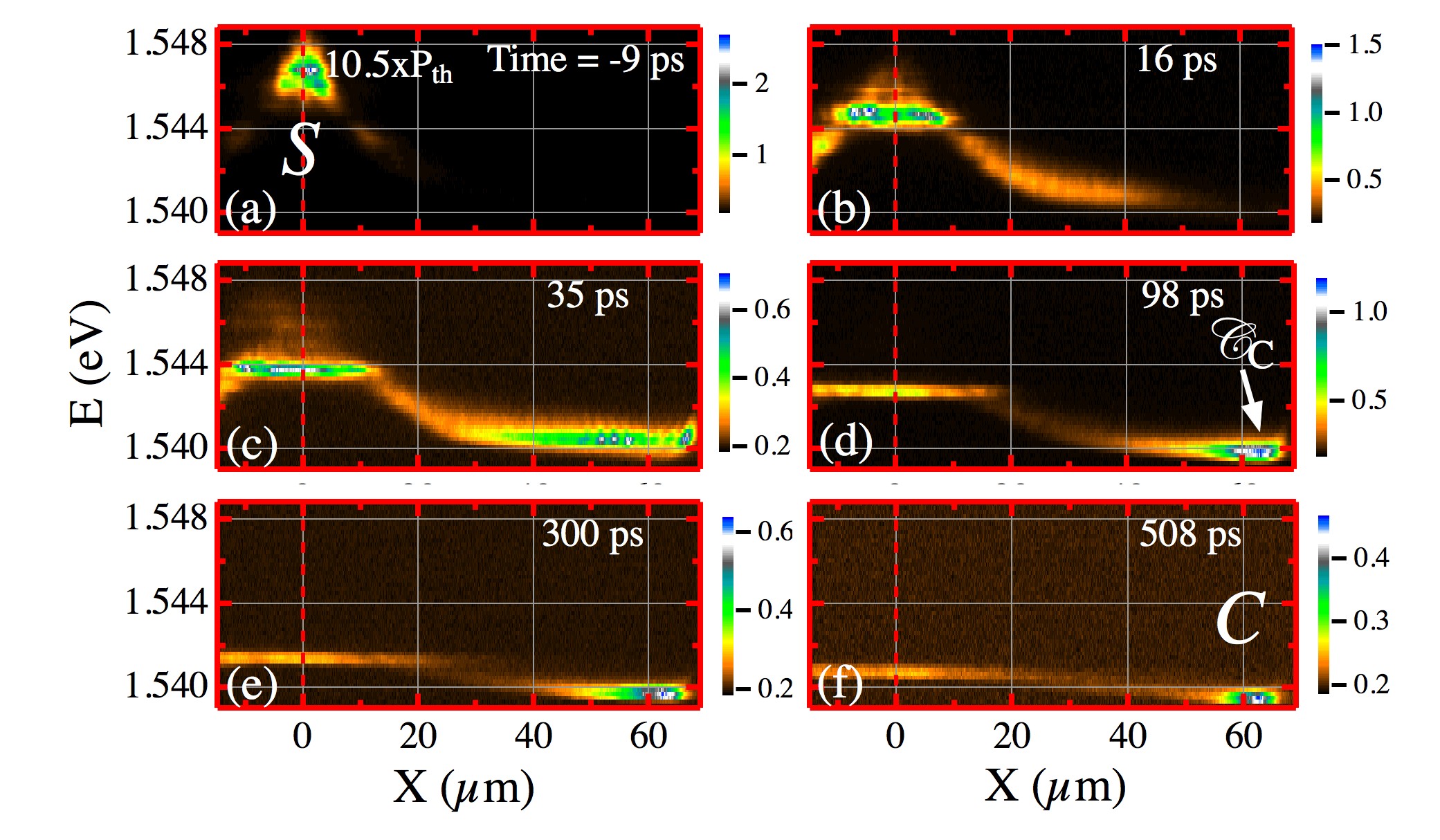}
\end{center}
\caption{(Color online) Energy vs. real space ($X$) for $P_S=10.5 \times P_{th}$ at different times shown by the labels. \emph{S}, \emph{C} and $\mathscr{C}_{C}$ mark the source, the collector and the trapped condensate positions, respectively. The intensity is coded in a false color scale shown on the right of each panel.}
\label{fig:fig4}
\end{figure}

In this section we present three time-resolved cases for different $P_S$ pump powers, Figs. ~\ref{fig:fig2}-\ref{fig:fig4}. Figure~\ref{fig:fig2} shows the dynamics of the emission when $P_S=P_{th}$. For each panel the time is displayed at the right upper corner, being the temporal origin set at the instant when the \emph{S} intensity is maximum. Panel (a) shows that the emission from \emph{S} at -24 ps occurs at 1.545 eV; at 31 ps, (b), the emission redshifts and a small spatial expansion around 0~$\mu$m is observed; propagating polaritons, expanding more rapidly towards the border, at an energy of1.542 eV, are detected at 194 ps (c), eventually reflecting backwards, interfering coherently and creating the $\mathscr{C}_{C}$ condensate, weakly emitting at 1.539 eV at later times, 483 ps (d). Since the pump power is at threshold, the emission intensity of polaritons is weak and slightly higher than the noise level in all panels of Fig. ~\ref{fig:fig2}.

Let us note that at early times, the emission observed in Fig.~\ref{fig:fig2} appears blueshifted from the lower polariton minimum by an amount comparable to one-half of the Rabi splitting. This suggests that the emission at the source comes from polaritons with a strong excitonic character. For this reason we will refer to the emission from the source as arising from excitons, although the decrease in the blueshift over time corresponds to a continuous transition from excitonic polariton states to those with roughly equal excitonic and photonic fractions. The emission from the propagating states and collector region, at lower energy, is clearly a polaritonic emission.

It is also important to note that the duration over which the condensate is present greatly exceeds the polariton lifetime. This is because the condensate is continuously fed by high energy excitons (not visible in the spectrum and distinct from the excitonic states emitting at the source) excited by the non-resonant pulse. The emission at \emph{S} is determined by repulsive interactions with hot excitons which contribute a blueshift to the potential energy. As hot excitons decay from the system, either through recombination or condensation, this potential energy decreases over time. Once the polariton condensate has formed, due to the low density of polaritons, there are minimal energy relaxation processes, such that the propagating polaritons tend to conserve their energy as they spread out from the source (see Fig.~\ref{fig:fig2} (d)).

The dynamics of the emission increasing $P_S$ to $1.7 \times P_{th}$ is shown in Fig.~\ref{fig:fig3}: the initial excitonic emission at the source takes place at 1.546 eV (a), slightly higher than before, due to increased blueshift due to a larger hot-exciton repulsion. At $t=73$ ps (b), an essential difference with respect to the case of Fig.~\ref{fig:fig2} (b) is revealed: polaritons emit from a lower energy than that of the source, which is $\backsim$2 meV blueshifted; this situation holds during the first $\backsim$200 ps of the decay process. Panel (c) shows the arrival of polaritons at the ridge border at 140 ps, and the eventual condensation of $\mathscr{C}_{C}$ (d). This final relaxation phase in the dynamics takes place into a state defined in a minimum of the wire structural potential located at the wire edge. A clear indication of the polariton coherence is evidenced by the interference at 1.540 eV between counter-propagating wave-packets. The source population at \emph{S}, still 1 meV blueshifted with respect to propagating polaritons, expands around $X=0$ as it decays in energy, and continuously feeds the propagating polariton condensate, increasing its effective lifetime (e). Finally, as shown in (f) at 617 ps, polaritons at \emph{S} merge with those propagating along the ridge. The emission is still observed for times as large as $\backsim1$~ns (not shown).

The case of the highest source power used in our experiments is shown in Fig.~\ref{fig:fig4}: at $-9$ ps the excitonic source population emits at 1.547 eV (a). The progressive spatial expansion of the excitonic population and the fast relaxation of the polariton condensate, as it propagates towards the right side, at 1.541 eV, reaching the ridge edge at 35 ps, is shown in panels (b) and (c).$\mathscr{C}_{C}$ is now slightly blueshifted, with respect to its energy at lower $P_S$ conditions, to 1.540 eV, due to the higher density condensate population at this place of the ridge (d). At later times, as those shown in (e,f) for 300 and 508 ps, the population at \emph{S} decreases and expands in space whilst $\mathscr{C}_{C}$ redshifts its energy emission due to its reduced occupancy.

Movies corresponding to Figs.~\ref{fig:fig2}-\ref{fig:fig4} are provided as Supplementary material.~\cite{suppmat}

\subsection{Two-beam excitation}
\label{subsec:two_beam}

\begin{figure}
\setlength{\abovecaptionskip}{-5pt}
\setlength{\belowcaptionskip}{-2pt}
\begin{center}
\includegraphics[trim=1.2cm 0.50cm 1.0cm 0.4cm, clip=true,width=1.0\linewidth,angle=0]{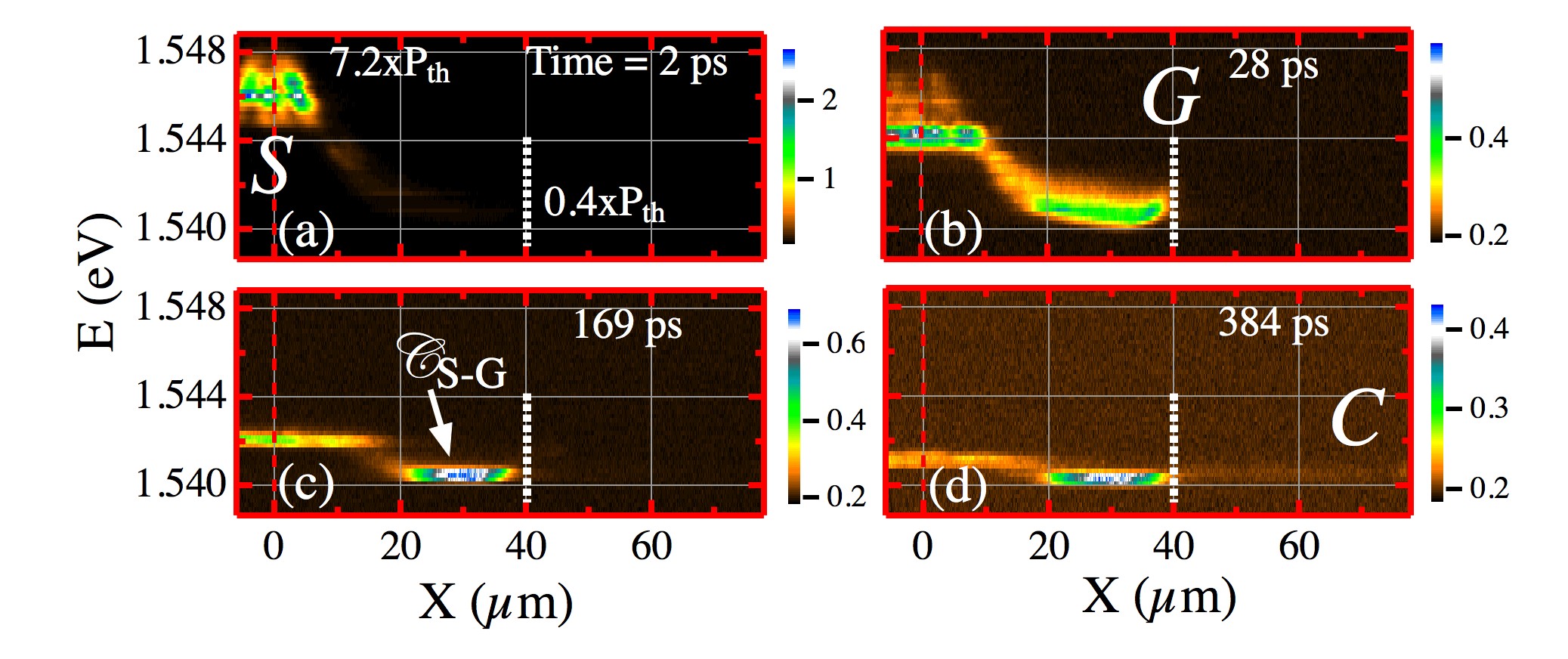}
\end{center}
\caption{(Color online) Energy vs. real space ($X$) for $P_S=7.2 \times P_{th}$ and $P_G=0.4\times P_{th}$ at different times shown by the labels. \emph{S}, \emph{G}, \emph{C} and $\mathscr{C}_{S-G}$ mark the source, the gate, the collector and the, between \emph{S-G}, trapped condensate positions, respectively. The intensity is coded in a false color scale shown on the right of each panel.}
\label{fig:fig5}
\end{figure}

The introduction of a new secondary pulse, dubbed before as gate (\emph{G}), between \emph{S} and \emph{C}, adds new interaction phenomena. The existence of two condensates becomes very clear in this case: one of them located initially between \emph{S}-\emph{G}, $\mathscr{C}_{S-G}$, which eventually becomes propagating, and a second one, already labelled as $\mathscr{C}_{C}$. The polariton propagation towards \emph{C} along the ridge can be hindered with a below-threshold intensity gate beam, see Fig.~\ref{fig:fig1} (b) and Fig.~\ref{fig:fig5}, rendering the $\mathscr{C}_{C}$ switch-off and creating the trapped condensate $\mathscr{C}_{S-G}$. As the \emph{G}-repulsive potential gradually decreases in time, a tiny fraction of $\mathscr{C}_{S-G}$ is able to tunnel through the barrier and it spreads between 40 and 80~$\mu$m, see Fig.~\ref{fig:fig5} (d). This configuration has been already discussed in detail in Ref.~\onlinecite{anton:261116}.

\begin{figure}
\setlength{\abovecaptionskip}{-5pt}
\setlength{\belowcaptionskip}{-2pt}
\begin{center}
\includegraphics[trim=1.2cm 0.6cm 0.7cm 0.4cm, clip=true,width=1.0\linewidth,angle=0]{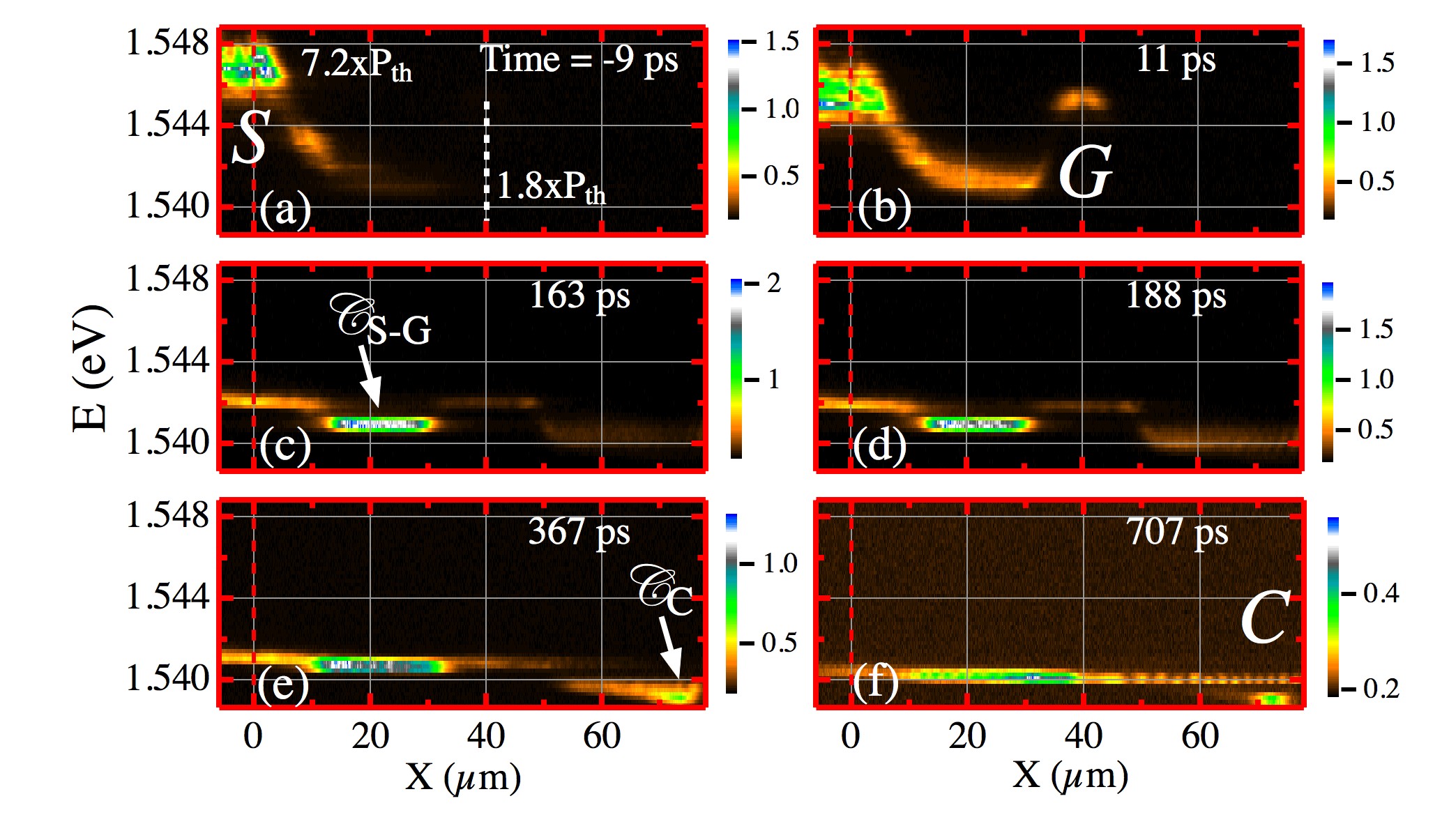}
\end{center}
\caption{(Color online) Energy vs. real space ($X$) for $P_S=7.2 \times P_{th}$ and $P_G=1.8\times P_{th}$ at different times shown by the labels. \emph{S}, \emph{G}, \emph{C}, $\mathscr{C}_{S-G}$ and $\mathscr{C}_{C}$ mark the source, the gate, the collector, the trapped condensate between \emph{S-G} and the trapped one at \emph{C} positions, respectively. The intensity is coded in a false color scale shown on the right of each panel.}
\label{fig:fig6}
\end{figure}

Figure~\ref{fig:fig6} shows the dynamics for $P_S=7.2\times P_{th}$ and $P_G=1.8\times P_{th}$. The visibility of the emission at \emph{G} is delayed by $\backsim$20 ps with respect to that at \emph{S}, despite of the fact that both beams reach the sample simultaneously, panels (a,b). This delay is due to the power dependence of the emission rise time, which increases with decreasing power. The $\mathscr{C}_{S-G}$ condensate lies at a constant energy, 1.541 eV, remaining trapped, (c). On their own account, the emission energy of \emph{S} and \emph{G} decay, at a rate determined by the carrier density and the carrier-carrier interactions, until they reach 1.542 eV for both. The population between \emph{G} and \emph{C}, mainly created by the \emph{G} pulse, propagates towards the border (c), being reflected (d) and forming the $\mathscr{C}_{C}$ condensate (e). When the \emph{G} barrier further decays the $\mathscr{C}_{S-G}$ condensate becomes propagating and coherent interference patterns are generated from counter-propagation, see panel (f).

\begin{figure}
\setlength{\abovecaptionskip}{-5pt}
\setlength{\belowcaptionskip}{-2pt}
\begin{center}
\includegraphics[trim=1.2cm 0.6cm 0.7cm 0.4cm, clip=true,width=1.0\linewidth,angle=0]{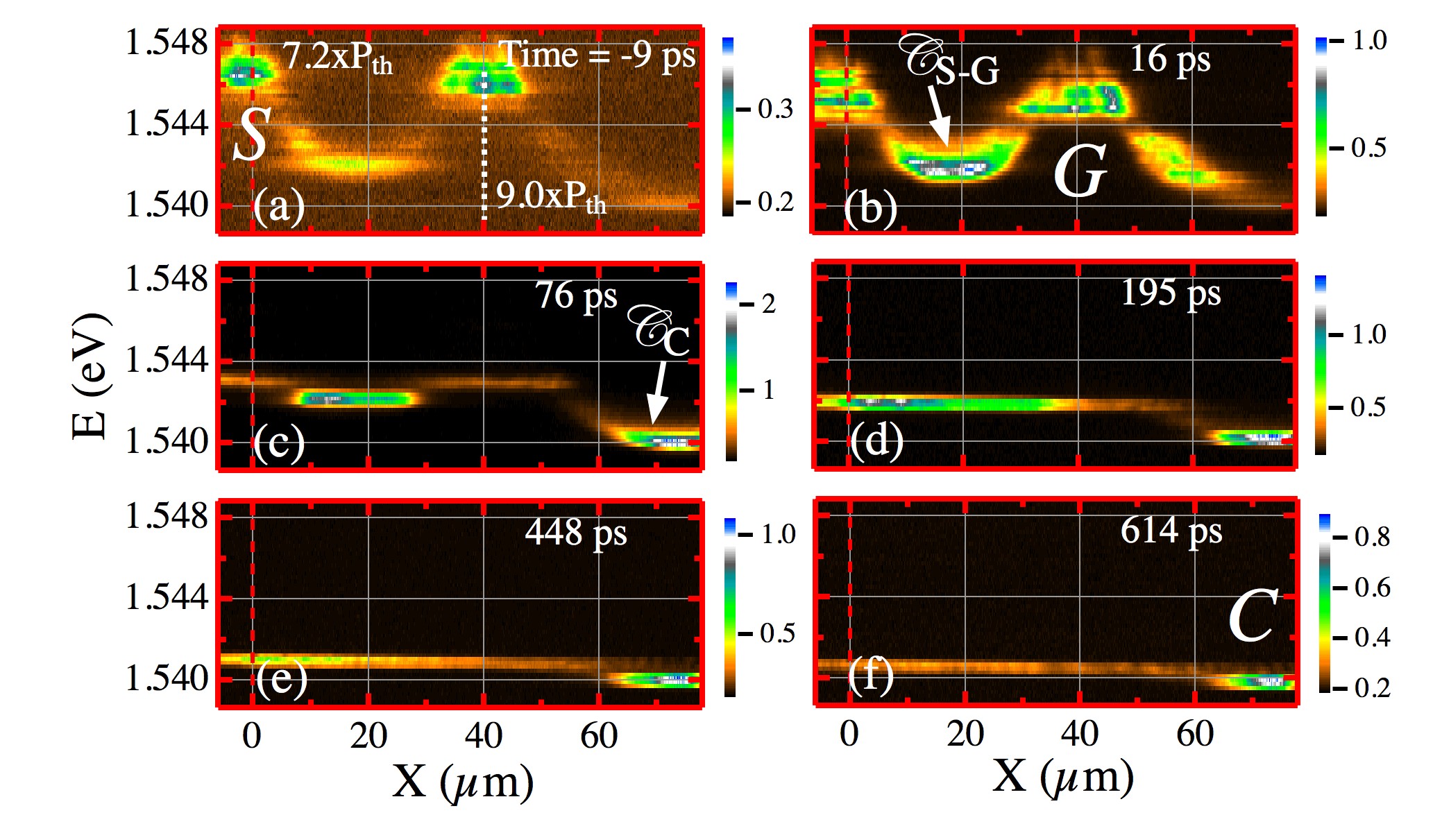}
\end{center}
\caption{(Color online) Energy vs. real space ($X$) for $P_S=7.2 \times P_{th}$ and $P_G=9.0\times P_{th}$ at different times shown by the labels. \emph{S}, \emph{G}, \emph{C}, $\mathscr{C}_{S-G}$ and $\mathscr{C}_{C}$ mark the source, the gate, the collector, the trapped condensate between \emph{S-G} and the trapped one at \emph{C} positions, respectively. The intensity is coded in a false color scale shown on the right of each panel.}
\label{fig:fig7}
\end{figure}

For completeness, Fig.~\ref{fig:fig7} depicts the case corresponding to large values of $P_G$. Panel (a) depicts the excitonic emission at 1.547 eV when the laser beams arrive at \emph{S} and \emph{G}. $\mathscr{C}_{S-G}$ is trapped around $X$=20~$\mu$m, blueshifted up to 1.542 eV, due to repulsive interactions (b), whilst polaritons between \emph{G} and \emph{C} propagate towards the border. At 76 ps, a new condensate, $\mathscr{C}_{C}$, becomes trapped at 1.540 eV, and the emission energy of \emph{S} and \emph{G} reaches that of $\mathscr{C}_{S-G}$ (c). Due to the barrier reduction at \emph{G}, $\mathscr{C}_{S-G}$ propagates along the ridge from 0 to 60~$\mu$m (d). $\mathscr{C}_{C}$ remains confined for later times at a constant energy, whereas $\mathscr{C}_{S-G}$ decays and interferes with itself, panels (e,f).

Movies corresponding to Figs.~\ref{fig:fig5}-\ref{fig:fig7} are provided as Supplementary material.~\cite{suppmat}

\subsection{Power dependance of the energy/intensity decays}
\label{subsec:power}

In our sample, the emission above 1.544 eV is coming from excitonic states. The polariton emission lies at lower energies, down to 1.538 eV at the collector region. In this section, we analyze the dynamics of the energy and population relaxation along the full region of propagation of the condensates between \emph{S} and \emph{C} both in the presence or absence of \emph{G}, obtaining quantitative values for the energy time-decays and the optimal working conditions for the ON-state.

\begin{figure}
\setlength{\abovecaptionskip}{-5pt}
\setlength{\belowcaptionskip}{-2pt}
\begin{center}
\includegraphics[trim=0.8cm 0.1cm 1.9cm 0.7cm, clip=true,width=1.0\linewidth,angle=0]{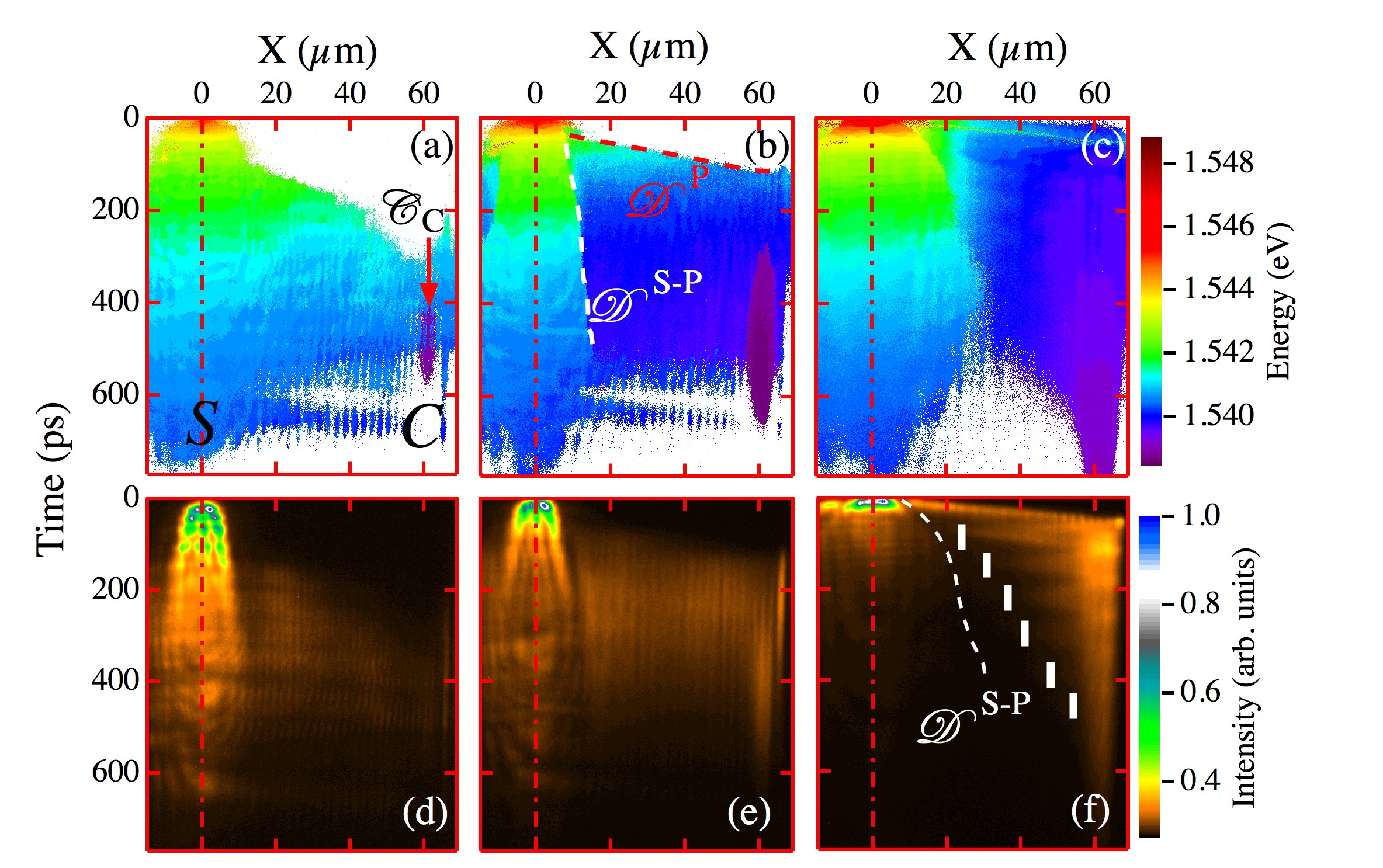}
\end{center}
\caption{(Color online) (a-c) Energy/(d-f) Intensity of the emission vs. real space ($X$) and time for different source excitation powers: (a/d) 1.0$\times P_{th}$ (b/e) 1.7$\times P_{th}$ and (c/f) 10.5$\times P_{th}$. \emph{S}, \emph{C} and $\mathscr{C}_{C}$ mark the source, the collector and the trapped condensate at \emph{C} positions, respectively. The $\mathscr{D}^{P}$ ($\mathscr{D}^{S-P}$) line marks the horizon between propagating polaritons and noise (carriers at \emph{S} and propagating polaritons), see text for further details. The information is coded in a linear/logarithmic false color scale shown on the right side of the upper/lower row.}
\label{fig:fig8}
\end{figure}

\begin{figure}
\setlength{\abovecaptionskip}{-5pt}
\setlength{\belowcaptionskip}{-2pt}
\begin{center}
\includegraphics[trim=0.8cm 0.1cm 1.9cm 0.7cm, clip=true,width=1.0\linewidth,angle=0]{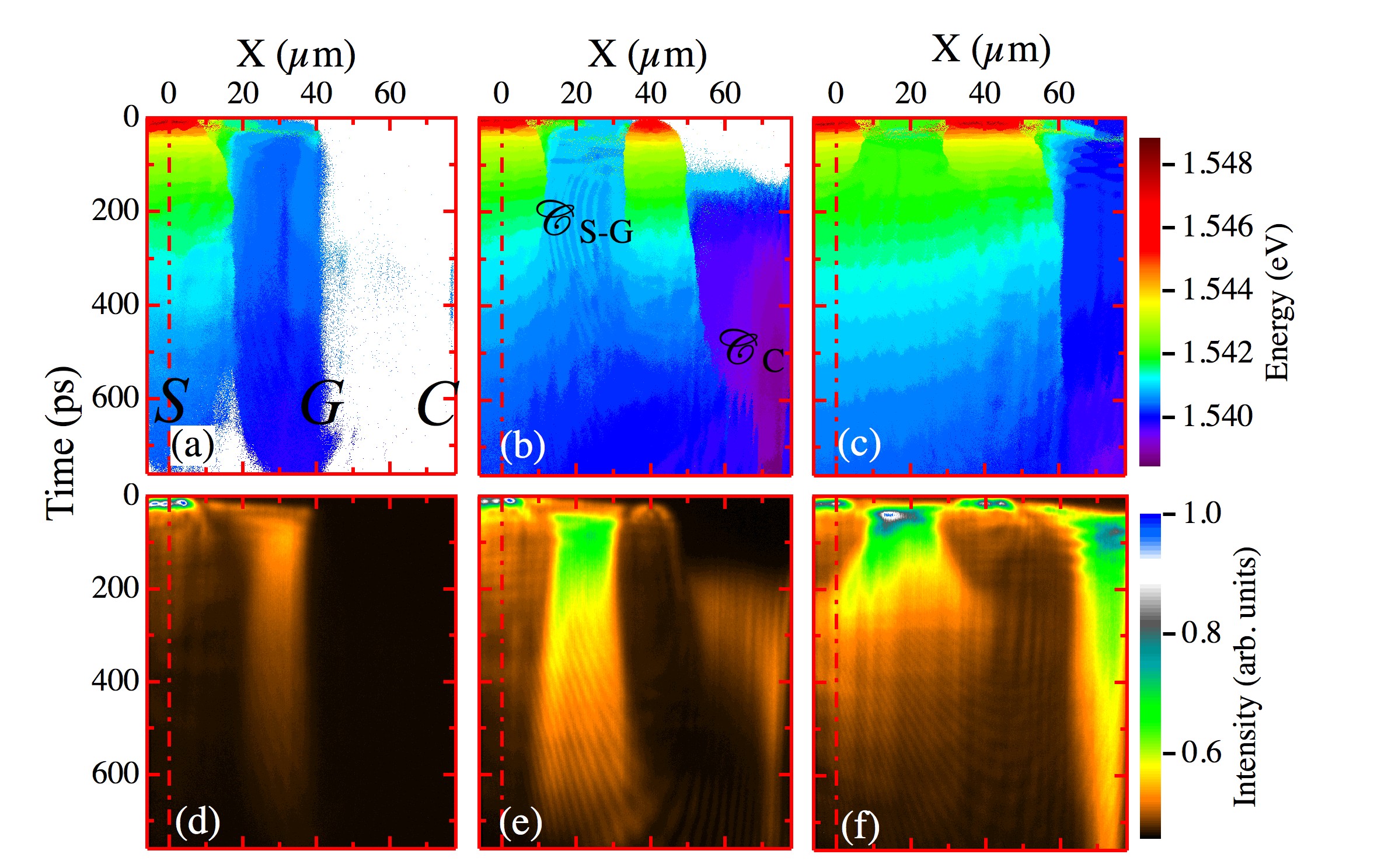}
\end{center}
\caption{(Color online) (a-c) Energy/(d-f) Intensity of the emission vs. real space ($X$) and time for a constant source excitation power $P_S=7.2\times P_{th}$ and different gate powers $P_G$: (a/d) 0.4$\times P_{th}$ (b/e) 1.8$\times P_{th}$ and (c/f) 9.0$\times P_{th}$. \emph{S}, \emph{G}, \emph{C}, $\mathscr{C}_{S-G}$ and $\mathscr{C}_{C}$ mark the source, the gate, the collector, the trapped condensate between \emph{S-G} and the trapped condensate at \emph{C} positions, respectively, see text for further details. The information is coded in a linear/logarithmic false color scale shown on the right side of the upper/lower row.}
\label{fig:fig9}
\end{figure}

Figures~\ref{fig:fig8} and \ref{fig:fig9} show spatial-temporal maps of the energy (a-c)/intensity (d-f) evolution of the emission for the same power values as those used in Figs.~\ref{fig:fig2}-\ref{fig:fig4} and \ref{fig:fig5}-\ref{fig:fig7}, respectively. Panels (a-c) have been obtained identifying the time at which the maximum emission intensity takes place, at a given \emph{X} position on the ridge, for every energy: this gives a point in the map whose energy is coded with the false-color scale shown on the right-hand side of the upper row. Note that all the information concerning the strength of the emission, and therefore the polariton population, is lost in this representation.~\cite{energynote} The complementary information is encoded in the second row in Figs.~\ref{fig:fig8} and \ref{fig:fig9}, giving in this case the polariton population from integrating all emission energies. These plots provide a straight and precise insight on the energy/intensity decay of the population at every position along the ridge.

Let us start by considering the one-beam excitation compiled in Fig.~\ref{fig:fig8}. In panels (a,b) the energy trap created at $\backsim60~\mu$m due to the potential discontinuity close to the border of the ridge is clearly observed: $\mathscr{C}_{C}$, emitting at 1.539 eV, is separated by a small energy gap from the polaritons propagating above at 1.540 eV. The horizon, $\mathscr{D}^{P}$, on the right upper corners, between white and colored points is given by the arrival of polaritons at different positions along the ridge. Another discontinuity is observed between the decay of carriers at \emph{S} and the propagating polaritons, $\mathscr{D}^{S-P}$. The power dependence of both discontinuities is evident in these panels and gives information about the speed of propagation of different emitting species. At $P_{th}$, the border, $\mathscr{D}^{S-P}$, between carriers at \emph{S} and polaritons, whose propagation is seen for $X\gtrapprox15~\mu$m, is absent (a), because the energy of excitons and polaritons decay at the same rate, but it becomes very clear in panels (b,c). The speed of propagation of the carriers can be obtained from the slope of $\mathscr{D}^{S-P}$ and $\mathscr{D}^{P}$ lines. For the carriers at \emph{S} in panel (b), $\mathscr{D}^{S-P}$ is almost straight, therefore a mean speed value, $v^S \scriptstyle \left(@1.7\times P_{th}\right)$, can be obtained amounting to $\backsim 0.02~\mu$m/ps. At the highest power (c), $v^S \scriptstyle \left(@10.5\times P_{th}\right)$ initially has increased by a factor of $\backsim$3 as compared to $v^S \scriptstyle \left(@1.7\times P_{th}\right)$, but the strong non-linearities associated with the high carrier densities lead to the appearance of deceleration rendering a gradual decrease of $v^S$. The spatial extension of the carriers around \emph{S} also widens with increasing power, almost doubling its value from $\backsim$30 to $\backsim$60$~\mu$m at 400 ps as seen in (b) and (c) panels. It is also noticeable that the energy decay of the carriers is spatially flat in the region enclosed by $\mathscr{D}^{S-P}$.

The acceleration/deceleration of the propagating polaritons is distinct in the slope changes of $\mathscr{D}^{P}$, panels (a-c). For $P_S=P_{th}$, a rough estimation of the speed obtains $v^P \scriptstyle \left(@P_{th}\right)$$=0.4~\mu$m/ps; $v^P$ increases to $0.6(1)$ and $1.1(1)~\mu$m/ps for $P_S=1.7\times P_{th}$ and $10.5\times P_{th}$, respectively. In the later case $v^P$ amounts to $0.3~\%$ of the speed of light in vacuum. The formation of $\mathscr{C}_{C}$ at threshold, (a), is seen by the purple (1.539 eV) oval shape at ($\backsim60~\mu$m, 400-600 ps). The enhancement of $v^P$ together with that of stimulated scattering processes with power give rise to an earlier appearance of $\mathscr{C}_{C}$ at $\backsim$280 ps lasting for 400 ps, almost doubling its spatial extent, at $P_S=1.7 \times P_{th}$ (b). The values for $v^P$ are in agreement with others reported in the literature (see, for example, Ref. \onlinecite{Amo2009}). The much smaller values for $v^S$ are due to the larger exciton mass compared to that of polaritons. The energy gap between $\mathscr{C}_{C}$ and the propagating polaritons dissolves at $P_S=10.5 \times P_{th}$ due to the very large number of polaritons and the very fast formation of this condensate. Finally, let us remark that the ballistic propagation of polaritons is evidenced in panel (b) by the constant energy (same color), for a given time, seen in the region enclosed by the $\mathscr{D}^{S-P}$-border and \emph{C}. However, in case (c) a gradual change in energy (color) is observed, indicating the energy loss during the polariton propagation towards \emph{C}.

We briefly discuss now the density maps for different power excitation shown in the lower row of Fig.~\ref{fig:fig8}, in a normalized, logarithmic false-color scale shown on their right-hand side. Panels (d-f) show that the main emission intensity arises from the population at \emph{S}, with a gradual expansion towards \emph{C} with a much lower polariton population. The emission-intensity decay becomes faster with increasing $P_S$ power. For $P_S=1.7 \times P_{th}$, (b), an enhanced emission following the $\mathscr{D}^{S-P}$ is apparent; interferences of polaritons in the region between $\backsim$20 and $\backsim$60~$\mu$m are visible; the formation of $\mathscr{C}_{C}$ appears at 280 ps. Panel (f) shows several reflections of condensed polaritons between the ridge edge and the left-bouncing positions marked with white bars, which are determined by the potential delimited by $\mathscr{D}^{S-P}$ and the energy of the bouncing condensates: the longer the time, the larger the energy loss of the polaritons, which become less able to climb the barrier side, as borne out by the progressively increasing distance between the bars and the $\mathscr{D}^{S-P}$ line, obtained from panel (c) and depicted with a white dotted line. At $\backsim$60~$\mu$m and 100 ps a considerable amount of population forms the $\mathscr{C}_{C}$ condensate.

We turn now to the two-beam excitation compiled in Fig.~\ref{fig:fig9}. Panel (a) displays the energy decay of the polaritons in the OFF-state for $P_S=7.2\times P_{th}$ and $P_G=0.4\times P_{th}$. The $\mathscr{C}_{S-G}$ condensate, extending 20~$\mu$m, reveals an almost constant energy emission in time. The contrast of the OFF-state is high as assessed by the negligible amount of polaritons that goes through the \emph{G} potential (d); only a hint of the polaritons that were able to tunnel through is seen at (80~$\mu$m, 400 ps) in panel (a), that codifies the energy but not the intensity of the signal. The ratio $I\left( \mathscr{C}_{S-G}\right)/I\left(S\right)$is much larger than $I\left( \mathscr{C}_{C}\right)/I\left(S\right)$, obtained in the one beam case since $\mathscr{C}_{S-G}$ is trapped closer to \emph{S} and its feeding process is more efficient. Increasing $P_G$ to $1.8 \times P_{th}$ both \emph{S} and \emph{G} beams contribute to the formation and trapping of polariton condensates (b), $\mathscr{C}_{S-G}$ and $\mathscr{C}_{C}$. Figure~\ref{fig:fig9} (c), for $P_G=9.0 \times P_{th}$, shows that, for the first $\backsim100$ ps, the energy decays at \emph{S} and \emph{G} are much faster than those of the polariton condensates. For longer times, $t\geq100$ ps, the energy decay of the populations at \emph{S}, $\mathscr{C}_{S-G}$ and \emph{G} is almost identical; however, $\mathscr{C}_{C}$ is always at a lower energy due to the trapping at \emph{C}.

A further inspection of the energy-integrated intensity maps shows that in panel (e), at 300 ps, when the \emph{G}-barrier has considerably decayed, so that its energy coincides with that of $\mathscr{C}_{S-G}$, the $\mathscr{C}_{S-G}$ condensate starts expanding along the ridge; concomitantly a slanted interference pattern is obtained, revealing the dynamics of merging counter-propagating polaritons.

The $\mathscr{C}_{S-G}$ formation time, $\backsim$70 ps, is much shorter than of $\mathscr{C}_{C}$, $\backsim$350 ps, due to the fact that $P_S$ is much larger than $P_G$ and that both beams contribute to feed $\mathscr{C}_{S-G}$ while only the population at \emph{G} refills the $\mathscr{C}_{C}$ condensate, which reaches its maximum intensity emission at 400 ps. In panel (f), the high \emph{S}- and \emph{G}-pump powers make the $\mathscr{C}_{S-G}$ condensate very intense at 40 ps. The confluence of the \emph{S}- and \emph{G}-population with $\mathscr{C}_{S-G}$ takes place at 100 ps and 1.542 eV. Then $\mathscr{C}_{S-G}$ doubles its spatial width, as observed by the spreading cone of polaritons extending 20~$\mu$m at 40 ps to 40~$\mu$m at 250 ps. A clear back and forth bouncing of the $\mathscr{C}_{C}$ condensate between the \emph{G}-barrier and the ridge edge is observed during the first 100 ps. After losing its kinetic energy at $t\backsim150$ ps, $\mathscr{C}_{C}$stops and emits for more than 600 ps, as its population is continuously fed by propagating polaritons at $\backsim1.541$ eV.

\begin{figure}
\setlength{\abovecaptionskip}{-5pt}
\setlength{\belowcaptionskip}{-2pt}
\begin{center}
\includegraphics[trim=1.0cm 0.45cm 0.6cm 0.2cm, clip=true,width=1.0\linewidth,angle=0]{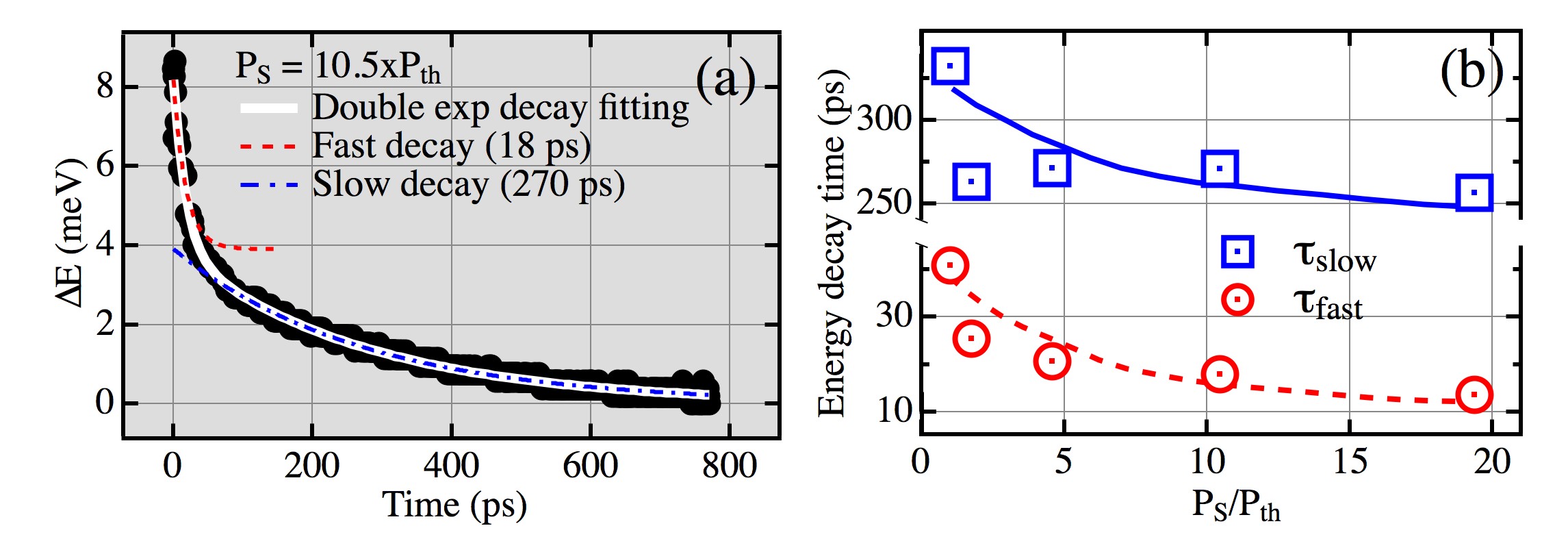}
\end{center}
\caption{(Color online) (a) Energy shift at \emph{S} vs. time for $P_S=10 \times P_{th}$ (black circles). Double exponential decay fit(solid white line): fast component ($\tau_{fast}=18$ ps, red dashed line), slow component ($\tau_{slow}=270$ ps, dot-dashed blue line). (b) $\tau_{slow}$/$\tau_{fast}$ decay times (blue square/red circle markers, blue full/ dashed red line is a guide to the eye) vs. normalized $P_S$ power.}
\label{fig:fig10}
\end{figure}

The energy maps shown in Figs.~\ref{fig:fig8}-\ref{fig:fig9} (a-c) allow to quantitatively analyze the energy decay at every $X$-position; in particular we present in Fig.~\ref{fig:fig10} this decay at the \emph{S} position for $P_S=10.5\times P_{th}$. The solid white line in Fig.~\ref{fig:fig10} (a) corresponds to its best fit to the sum of two exponentially decaying functions, shown separately by the dashed and dot-dashed lines. The double fashion decay is attributed to two different physical processes: a fast decay due to relaxation driven by exciton-exciton interactions and a slow one, attributed to the decreasing blueshift caused by the diminishing exciton and polariton populations. The rate of condensation can be expected to be faster at early times due to larger densities of carriers resulting in stronger stimulated scattering processes. A fast condensation rate results in an initial fast drop in the exciton density since excitons condense rapidly into polaritons that quickly decay. This drop in the exciton population gives a corresponding drop in the polariton population and so both blueshifts, due to polariton-exciton and polariton-polariton repulsion, drop sharply at early times. At longer times polariton condensation proceeds slower, due to weaker stimulated scattering and the exciton populations decay with a slow exponential dependence due to exciton recombination. Figure ~\ref{fig:fig10} (b) compiles the power dependence of the decay times: both decrease with increasing power, more markedly for $\tau_{fast}$ (circles), which decreases by $\backsim 65 \%$ for a 20 fold increase of power, whilst $\tau_{slow}$ (squares) only diminishes by $\backsim 20 \%$, revealing the larger influence of density in exciton-exciton scattering processes than in exciton-polariton ones.

\begin{figure}
\setlength{\abovecaptionskip}{-5pt}
\setlength{\belowcaptionskip}{-2pt}
\begin{center}
\includegraphics[trim=1.0cm 0.2cm 2.0cm 0.4cm, clip=true,width=1.0\linewidth,angle=0]{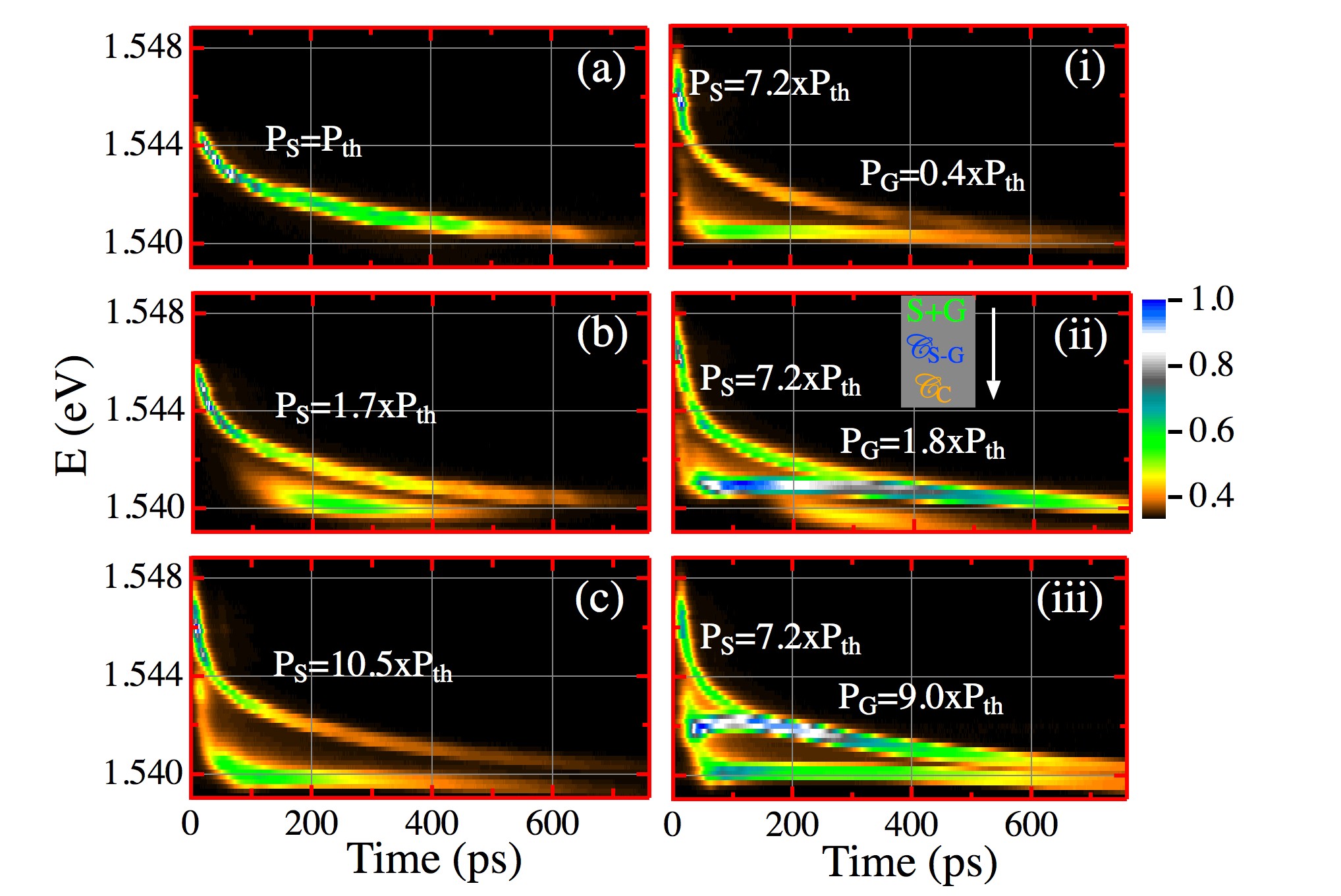}
\end{center}
\caption{(Color online) Energy decay, spatially-integrated, vs. time for one (a-c) and two beams (i-iii) excitation conditions. (a) $P_S= 1.0 \times P_{th}$, (b) 1.7$\times P_{th}$ and (c) 10.5$\times P_{th}$. For $P_S=7.2 \times P_{th}$: (i) $P_G=0.4\times P_{th}$, (ii) 1.8$\times P_{th}$ and (iii) 9.0$\times P_{th}$, see text for further details. The intensity is coded in a logarithmic false color scale shown on the right.}
\label{fig:fig11}
\end{figure}

The spatial integration of the data shown in Figs.~\ref{fig:fig2}-\ref{fig:fig7} reveals the total energy and intensity decay dynamics for the different configurations under study as shown in Fig.~\ref{fig:fig11}: panels (a-c)/(i-iii), correspond to one/two beam excitation under different $P_S$/$P_G$ powers. The addition of contributions from different population species gives rise to a very rich dynamics. Figures~\ref{fig:fig11} (a-c) exhibit a critical difference in the power dependence of the total decay: panel (a) shows a collective energy decay for \textbf{all} spatial positions along the ridge. Panels (b,c) show a low energy streak corresponding to a polariton condensate drop that propagates along the ridge with an almost constant energy, unveiling the ballistic propagation of the condensate. The two streaks presented in panel (i) correspond to the decay of population at \emph{S} (high energy one) and the emission of $\mathscr{C}_{S-G}$ for a typical switch OFF state (low energy one): the dynamics of both streaks is similar to those shown in panel (c), with the difference, not appreciated in the figure, that polaritons now are stopped just before the \emph{G} barrier.

The three traces appearing in panel (ii), ordered by decreasing energy, compile the emission from: the population at the source and the gate (\emph{S+G}), the $\mathscr{C}_{S-G}$ condensate and the $\mathscr{C}_{C}$ condensate, respectively. It is worthwhile noting the identical decay dynamics of the \emph{S} and \emph{G} populations, observed by the existence of only one streak for both populations. The $\mathscr{C}_{C}$ condensate shows an emission at $\backsim$1.539 eV, with a dynamics similar to that shown in panel (b). As the \emph{S} power is kept constant in this subset of experiments, the \emph{G} power permits manipulating on demand the amount of condensed polaritons at $\mathscr{C}_{S-G}$: if it would have been formed only by the \emph{S} pulse, its energy should decay slightly; however, the extra population injected by the \emph{G} pulse contributes with an additional blueshift giving rise to an increase of the $\mathscr{C}_{S-G}$ emission energy, hinted at $\backsim$ 300 ps in panel (ii), which becomes clearly visible in panel (iii). In this latter panel, the \emph{S+G} decays are also superimposed and $\mathscr{C}_{C}$ emits at a constant energy of 1.540 eV for $t > 50$ ps. It is important to note that in panels (ii,iii), the additional polaritons provided by the \emph{G} pulse make the $\mathscr{C}_{S-G}$ condensate the highest populated state in the device with an emission intensity even larger than that of $S+G$ together.

\begin{figure}
\setlength{\abovecaptionskip}{-5pt}
\setlength{\belowcaptionskip}{-2pt}
\begin{center}
\includegraphics[trim=1.0cm 0.2cm 0.5cm 0.4cm, clip=true,width=1.0\linewidth,angle=0]{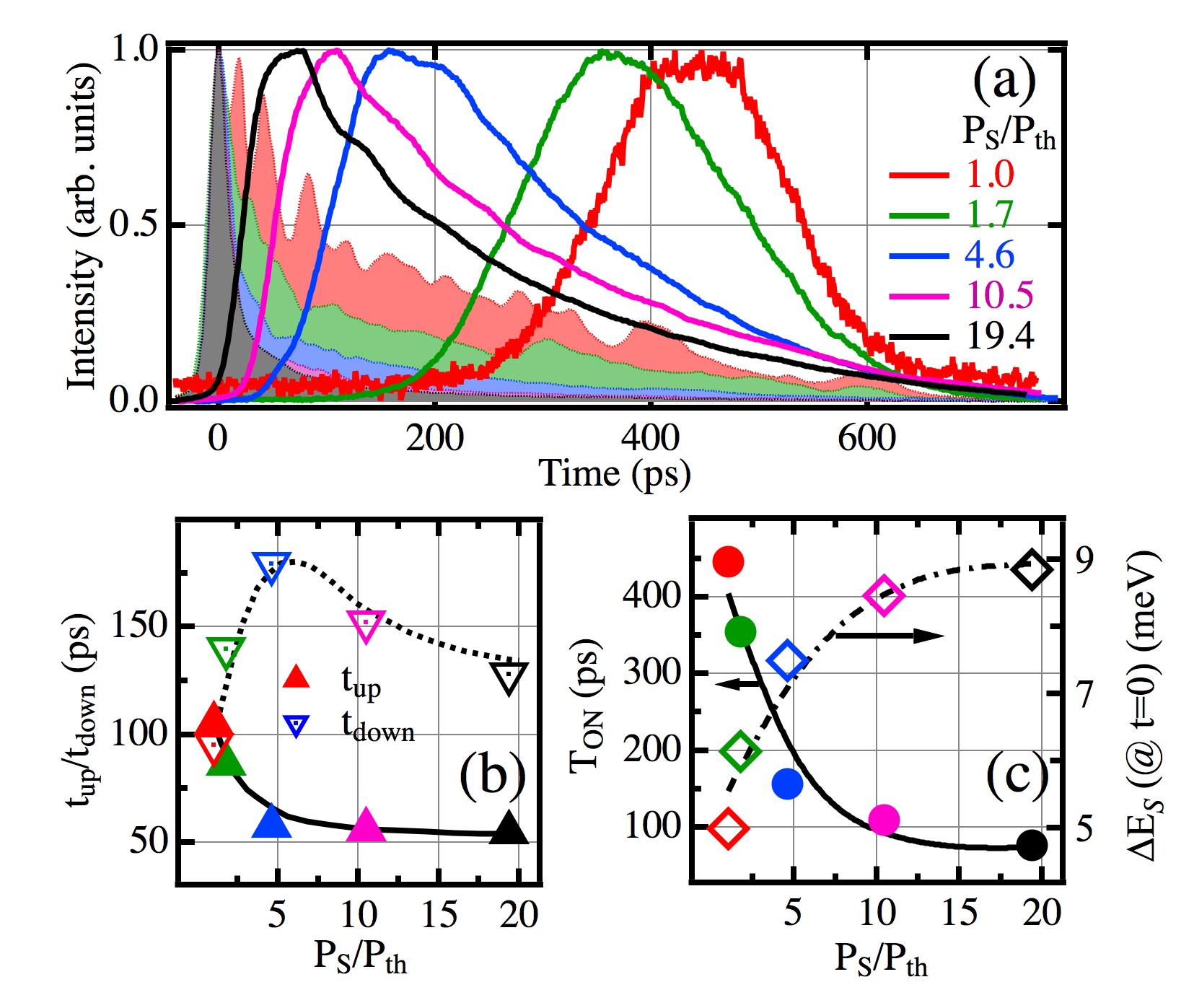}
\end{center}
\caption{(Color online) (a) Integrated intensity at \emph{S} position in filled lines and integrated intensity at \emph{C} in full line for $P_S=1.0,1.7,4.6,10.5,19.4 \times P_{th}$. (b) Up/ Down triangles show the raising/decreasing time $t_{up}$/$t_{down}$ from 0.5 to 1/1 to 0.5. (c) Full circles show the temporal separation between \emph{S} and \emph{C} intensity peaks; open diamonds depict the initial energy shift of the emission at \emph{S}. Same color legend for the values of $P_S$ are used in (a), (b) and (c).}
\label{fig:fig12}
\end{figure}

\subsubsection{Optimization of the switching time}
\label{subsec:opt_switch}

Let us consider the optimal power conditions for the ON state for a \emph{S-C} spatial separation of $\backsim60$~$\mu$m. We present in Fig.~\ref{fig:fig12} the main effects of the $P_S$ power on the transistor switch ON state, which was illustrated before in Figs.~\ref{fig:fig2}-\ref{fig:fig4}. Figure~\ref{fig:fig12} (a) plots the normalized intensity dynamics of the source, at $X=0$, (shadowed traces) and that of $\mathscr{C}_{C}$, at $X=60$~$\mu$m, (full lines), for different $P_S/P_{th}$ values. We define the switch ON time, $T_{ON}$, as the temporal delay between the \emph{S} maximum intensity and that of the $\mathscr{C}_{C}$. It is clearly observed that $T_{ON}$ decreases, and the shape of the $\mathscr{C}_{C}$ time-evolutions becomes more asymmetric with increasing $P_S$. The asymmetry of the $\mathscr{C}_{C}$ temporal evolution (see Fig.~\ref{fig:fig12} (a)), which strongly depend on $P_S$, is characterized in Fig.~\ref{fig:fig12} (b), where we define a raise time, $t_{up}$ (up triangles), given by the time spent to raise from an intensity of 0.5 up to the maximum value of 1. Similarly, $t_{down}$ (down triangles) is given by the time interval in which the intensity falls from a value of 1 to 0.5. A non monotonic dependence of $t_{down}$ on power is observed with a sharp raise at low $P_S$ values and a gradual fall for high ones: if the aim is to create a long lived ON state, the optimal power corresponds to $P_S/P_{th}\approx 7$, where $t_{down}\backsim175$ ps. On its own hand, the raise time, $t_{up}$, decreases monotonically with increasing $P_S$, reaching a minimum value of $\backsim50$ ps: a marked dependence at small powers, followed by an almost negligible decay at high ones, results in an optimum power to create a fast response transistor at similar powers than those required for a long lived ON state. Figure~\ref{fig:fig12} (c) shows the power dependence of $T_{ON}$ (full circles) together with the initial energy shift of the emission at \emph{S} ($\Delta E_S$, open diamonds). The monotonous decrease of $T_{ON}$ with power (increase of switching rate) is linked to the increase of $\Delta E_S$, due to the enhanced polariton acceleration from \emph{S} to \emph{C} produced by the augmented photo-generated repulsive excitonic potential, but other contributions as, for example, increase of stimulated scattering processes in the creation of $\mathscr{C}_{C}$ are also responsible for the quickening of $T_{ON}$. The minimum value of $T_{ON}$, $\backsim80$ ps, corresponds to a polariton propagation speed of $\backsim1.1$~$\mu$m/ps in agreement with the values of $v^P$ obtained from the horizon established by $\mathscr{D}^{P}$ in the energy maps of Fig~\ref{fig:fig8} (a-c). Finally, we should mention that our results indicate that the optimal conditions for gating are obtained for $P_G=0.6\times P_{th}$, in agreement with the previous results of Ref.~\onlinecite{Gao:2012tg}. At this power, the maximum attenuation of the $\mathscr{C}_{C}$ condensate is obtained yielding the highest contrast for the OFF state; at lower values of $P_G$ the traveling polaritons are not gated efficiently and at higher values the gate starts feeding $\mathscr{C}_{C}$.

\section{Model}
\label{sec:model}

To model our experimental results theoretically, we make use of a phenomenological treatment of polariton energy-relaxation processes taking place in the system. Such processes are not only responsible for the relaxation of hot excitons (injected by the pump) into polaritons in the form of a condensate, but also for the further relaxation in energy of the polariton condensate as it propagates. This latter energy relaxation process can be strongly influenced by a spatially dependent potential coming from repulsion from the hot excitons. Let us first introduce the description of the polariton condensate, which we will later couple to a description of higher energy excitons.

A fundamental feature of Bose-Einstein condensates is their spatial coherence that allows them to be well described with a mean-field approach.~\cite{Dalfovo1999} The Gross-Pitaevskii equation has been developed to describe the non-equilibrium dynamics of condensed polaritons, where losses due to the short polariton lifetime~\cite{carusotto04} and gain due to non-resonant pumping~\cite{Wouters2007,keeling08} were included phenomenologically. In such form, a variety of recent experiments can be modeled, including, for example, experiments on polariton transport~\cite{Wertz:2012ee}, spatial pattern formation~\cite{Manni:2011fu,Christmann2012} and spin textures.~\cite{Kammann2012} The Gross-Pitaevskii equation for the polariton wave-function, $\psi(x,t)$, is:
\begin{align}
i\hbar\frac{d\psi(x,t)}{dt}&=\left[\hat{E}_{LP}+\alpha|\psi(x,t)|^2+V(x,t)\right.\notag\\
&\hspace{10mm}\left.+i\hbar\left( rN_A(x,t)-\frac{\Gamma}{2}\right)\right]\psi(x,t)\notag\\
&\hspace{5mm}+i\hbar\mathfrak{R}\left[\psi(x,t)\right]\label{eq:GP}
\end{align}
Here $\hat{E}_{LP}$ represents the kinetic energy dispersion of polaritons, which at low wavevectors can be approximated as $\hat{E}_{LP}=-\hbar^2\hat{\nabla^2}/\left(2m\right)$, with $m$ the polariton effective mass. $\alpha$ represents the strength of polariton-polariton interactions. Being repulsive ($\alpha>0$), these interactions allow both a spatially dependent blueshift of the polariton condensate energy and energy-conserving scattering processes. Our analysis shows, however, that neither of these effects play a dominant role in our experiments. The effective potential acting on polaritons caused by repulsive interactions between polaritons and higher energy excitons is more significant, and is responsible for the blocking of polariton propagation in the presence of a gate pump. The effective potential $V(x,t)$ can be divided into a contribution from three different types of hot exciton states, which will be described shortly, as well as a static contribution due to the wire structural potential, $V_0(x)$:
\begin{align}
V(x,t)&=\hbar\left[g_R N_A(x,t)+g_I N_I(x,t)+g_D N_D(x,t)\right]\notag\\
&\hspace{5mm}+V_0(x)
\end{align}
$N_A$, $N_I$ and $N_D$ correspond to density distributions of ``active", ``inactive" and dark excitons, respectively, as described below.

Experimental characterization has revealed that the static potential, $V_0(x)$, is non-uniform along the wire and exhibits a slight dip in the potential near the wire edge. $g_R$, $g_I$ and $g_D$ define the strengths of interaction with the various hot exciton states.

$N_A$ represents the density distribution of an ``active'' hot exciton reservoir.~\cite{Lagoudakis2011,Manni2012} These excitons have the correct energy and momentum for direct stimulated scattering into the condensate and so appear as an incoherent pumping term in Eq.~(\ref{eq:GP}) with $r$ the condensation rate. To describe the dynamics of the system, it is important to note that not all excitons in the system are in this active form. In fact, the non-resonant pumping creates excitons with very high energy and they must first relax in energy before becoming active. We can thus identify an ``inactive'' reservoir of hot excitons that is excited by the non-resonant pump but not directly coupled to the condensate. The dynamics of the exciton densities are described by rate equations:
\begin{align}
\frac{dN_A(x,t)}{dt}&=-\left(\Gamma_A+r|\psi(x,t)|^2\right)N_A(x,t)\notag\\
&\hspace{10mm}+\left(t_R+t^\prime_RN_I(x,t)\right)N_I(x,t)\label{eq:NA}
\end{align}
\begin{align}
\frac{dN_I(x,t)}{dt}&=-\left(\Gamma_I+t_R+t^\prime_RN_I(x,t)+t_D\right)N_I(x,t)
\label{eq:NI}
\end{align}
\begin{align}
\frac{dN_D(x,t)}{dt}&=t_DN_I(x,t)-\Gamma_DN_D(x,t)\label{eq:ND}
\end{align}
When solving the equations we start from the initial condition $N_A(x,0)=0$, $N_D(x,0)=0$ and introduce a density proportional to the pump intensity profile in the inactive reservoir, $N_I(x,0)$. This represents an instantaneous injection by the non-resonant ultra-short pulse used in the experiment. The inactive reservoir is coupled by both linear and non-linear terms to the active reservoir, described by $t_R$ and $t_R^\prime$, respectively.

We also account for a linear coupling to a dark exciton reservoir, $N_D(x,t)$, described by coupling rate $t_D$. Dark excitons are long-lived states that are optically inactive yet can nevertheless be populated as high energy excitations from the non-resonant pump relax in energy. The dark excitons introduce a long-lived repulsive contribution to the effective polariton potential, $V(x,t)$, and are thus efficient at gating propagating polaritons at long-times. $\Gamma_A$, $\Gamma_I$ and $\Gamma_D$ describe the decay rates of each of the reservoirs.

The feeding of dark excitons from the inactive reservoir represents processes where higher energy electron-hole pairs relax in energy forming dark exciton states. We have neglected any further conversion between bright and dark excitons. Nonlinear conversion has been shown to generate oscillations between bright and dark excitons.~\cite{Vina:2004fk, Shelykh:2005fk} However, these processes require coherent excitation of exciton-polaritons near the dark exciton resonance. In our case, we do not expect accumulation of exciton-polaritons at such an energy. Furthermore, the fact that no oscillations in the polariton condensate density were observed, suggests that any coupling between bright and dark states is slower than the condensation rate.

It is worthwhile mentioning that we have also considered hot-exciton diffusion along the wire when solving Eqs.~(\ref{eq:NA}-\ref{eq:ND}), using typical exciton diffusion rates; however no noticeable effect on polariton dynamics was observed.

Returning to Eq.~(\ref{eq:GP}), the decay of polaritons is accounted for by the decay rate $\Gamma$. The final term in Eq.~(\ref{eq:GP}) accounts for energy relaxation processes of condensed polaritons. Polaritons are expected to condense at the source into the lowest energy state, where they have zero kinetic energy and potential energy given by $V(x,t)$ (and an additional blueshift due to polariton-polariton interactions). While this is the lowest energy state available at the source, one notes that the potential energy can be reduced if polaritons propagate away from the source ($V(x,t)$ decreases away from the source, where the reservoir densities are weaker). If polaritons were to conserve their energy, then they would convert this potential energy into kinetic energy as they move away from the source, accelerating down the potential gradient. However, the polariton kinetic energy can be lost as polaritons scatter with acoustic phonons~\cite{Tassone1997} or hot excitons.~\cite{Porras2002} Surface scattering could be also responsible for this loss; however since we have considered a phenomenological energy relaxation, the actual mechanism that causes that relaxation does not play a direct role in our calculations. 

Previous methods to introduce energy relaxation into a description of polariton condensates have been based on the introduction of an additional decay of particles depending on their energy~\cite{Read:2009uq, Wouters2010, Wouters:2012fk, Wertz:2012ee} (occasionally known as the Landau-Khalatnikov approach). The polariton number can be conserved in such a process via the introduction of an effective chemical potential.~\cite{Wouters2010,Wouters:2012fk} The energy relaxation term is:
\begin{equation}
\mathfrak{R}[\psi(x,t)]=-\left(\nu+\nu^\prime|\psi(x,t)|^2\right)\left(\hat{E}_\mathrm{LP}-\mu(x,t)\right)\psi(x,t),\label{eq:relax}
\end{equation}
where $\nu$ and $\nu^\prime$ are phenomenological parameters determining the strength of energy relaxation.~\cite{Read:2009uq,Wouters2010,Wouters:2012fk,Wertz:2012ee} We do not attempt here a microscopic derivation of the energy relaxation terms, we only note that we can expect some energy relaxation at low polariton densities (described by the parameter $\nu$) as well as a stimulated component of the relaxation proportional to the polariton density, $|\psi(x,t)|^2$ (described by the parameter $\nu^\prime$). Note that the energy relaxation rate is assumed proportional to the kinetic energy of polaritons; polaritons will relax in energy until they decay from the system or until their kinetic energy is zero (such that they have zero in-plane wave-vector).

The local effective chemical potential, $\mu(x,t)$, can be obtained from the condition:
\begin{equation}
\left.\frac{\partial\sqrt{n(x,t)}}{\partial t}\right|_\mathfrak{R}=0,
\end{equation}
where, $\psi(x,t)=\sqrt{n(x,t)}e^{i\theta(x,t)}$ for $n(x,t)$ real, and:
\begin{align}
\left.\frac{d\psi(x,t)}{dt}\right|_\mathfrak{R}&\equiv\mathfrak{R}[\psi(x,t)]\notag\\
&=\left.\frac{\partial\sqrt{n(x,t)}}{\partial t}\right|_\mathfrak{R}e^{i\theta(x,t)}\notag\\
&\hspace{10mm}+\left.i\sqrt{n(x,t)}e^{i\theta(x,t)}\frac{\partial\theta(x,t)}{\partial t}\right|_\mathfrak{R}.
\end{align}
$|_\mathfrak{R}$ denotes the components of the derivatives due to the term $\mathfrak{R}[\psi(x,t)]$ in Eq.~(\ref{eq:GP}). Note that other terms in Eq.~(\ref{eq:GP}), namely the pumping and loss terms, do not conserve the number of condensed polaritons.

Although it would be desirable to define a mean free path between scattering events, this is not trivial because the energy-relaxation rate is both energy and density dependent.

Equations~(\ref{eq:GP}-\ref{eq:ND}) were solved numerically for different initial density profiles $N_I(x,0)$, corresponding to the different source and gate configurations studied experimentally. We used the following parameters in the theory: $m=7.3\times10^{-5}m_e$ (obtained from fits to the dispersions measured in Ref.~\onlinecite{Gao:2012tg}; $m_e$ is the free electron mass), $\alpha=2.4\times10^{-3}$meV$\mu m^2$ (Ref.~\onlinecite{Kammann2012}), $\Gamma=1/18$ ps$^{-1}$, $\Gamma_A=0.01$ ps$^{-1}$, $\Gamma_I=\Gamma_D=10^{-3}$ ps$^{-1}$, $t_R=10^{-4}$ ps$^{-1}$, $t_D=2\times10^{-4}$ ps$^{-1}$, $t_R^\prime=10^{-4}$ ps$^{-1}\mu$m$^2$, $g_R=g_I=0.04$ ps$^{-1}\mu$m$^2$, $g_D=0.5$ ps$^{-1}\mu$m$^2$, $\hbar\nu=0.014$, $\hbar\nu^\prime=0.075\mu$m$^2$. 

\subsection{Simulations for the one-beam excitation}
\label{subsec:sim_one_beam}

\begin{figure}
\setlength{\abovecaptionskip}{-5pt}
\setlength{\belowcaptionskip}{-2pt}
\begin{center}
\includegraphics[trim=0.3cm 0.2cm 0.5cm 0.4cm, clip=true,width=1.0\linewidth,angle=0]{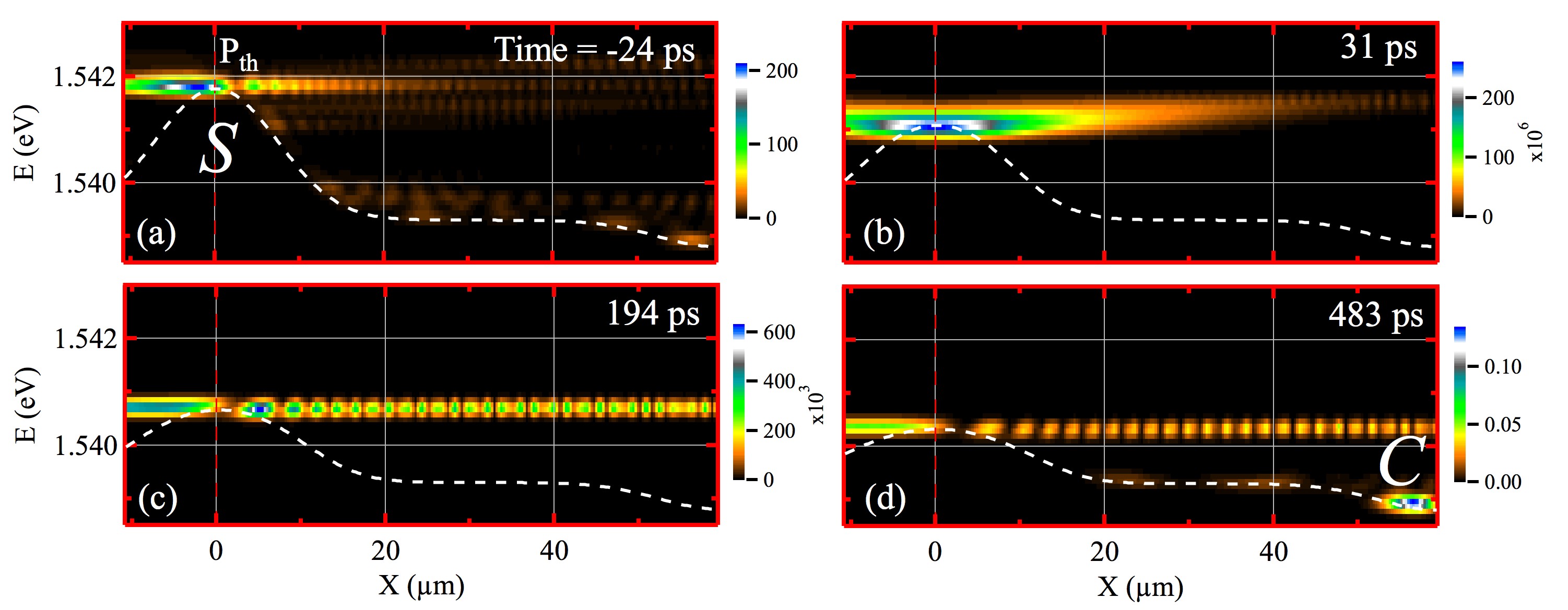}
\end{center}
\caption{(Color online) Energy vs. real space ($X$) for $P_S=P_{th}$ at different times shown by the labels. The white dashed curves show the evolution of the effective potential $V(x,t)$ due to hot exciton repulsion as well as the ridge structural potential (same for Figs.~\ref{fig:Time_2}-\ref{fig:Time_4}). \emph{S} and \emph{C} mark the source and the collector positions, respectively. The intensity is coded in a false color scale shown on the right of each panel.}
\label{fig:Time_1}
\end{figure}

Figure~\ref{fig:Time_1} shows the evolution of the spectrum in real space for $P_S=P_{th}$. As in the experimental case (Fig.~\ref{fig:fig2}), condensation initially takes place into a state blueshifted due to the repulsive interactions from the hot exciton reservoirs contributing to the effective potential $V(x,t)$. Over time, this blueshift decays resulting in progressively lower energy of the source state. In addition, a propagating polariton state can be observed, which forms interference fringes due to reflection from the end of the ridge. Energy relaxation is slow, appearing only at very long times due to the lack of stimulation by the low polariton density.

\begin{figure}
\setlength{\abovecaptionskip}{-5pt}
\setlength{\belowcaptionskip}{-2pt}
\begin{center}
\includegraphics[trim=0.3cm 0.2cm 0.5cm 0.3cm, clip=true,width=1.0\linewidth,angle=0]{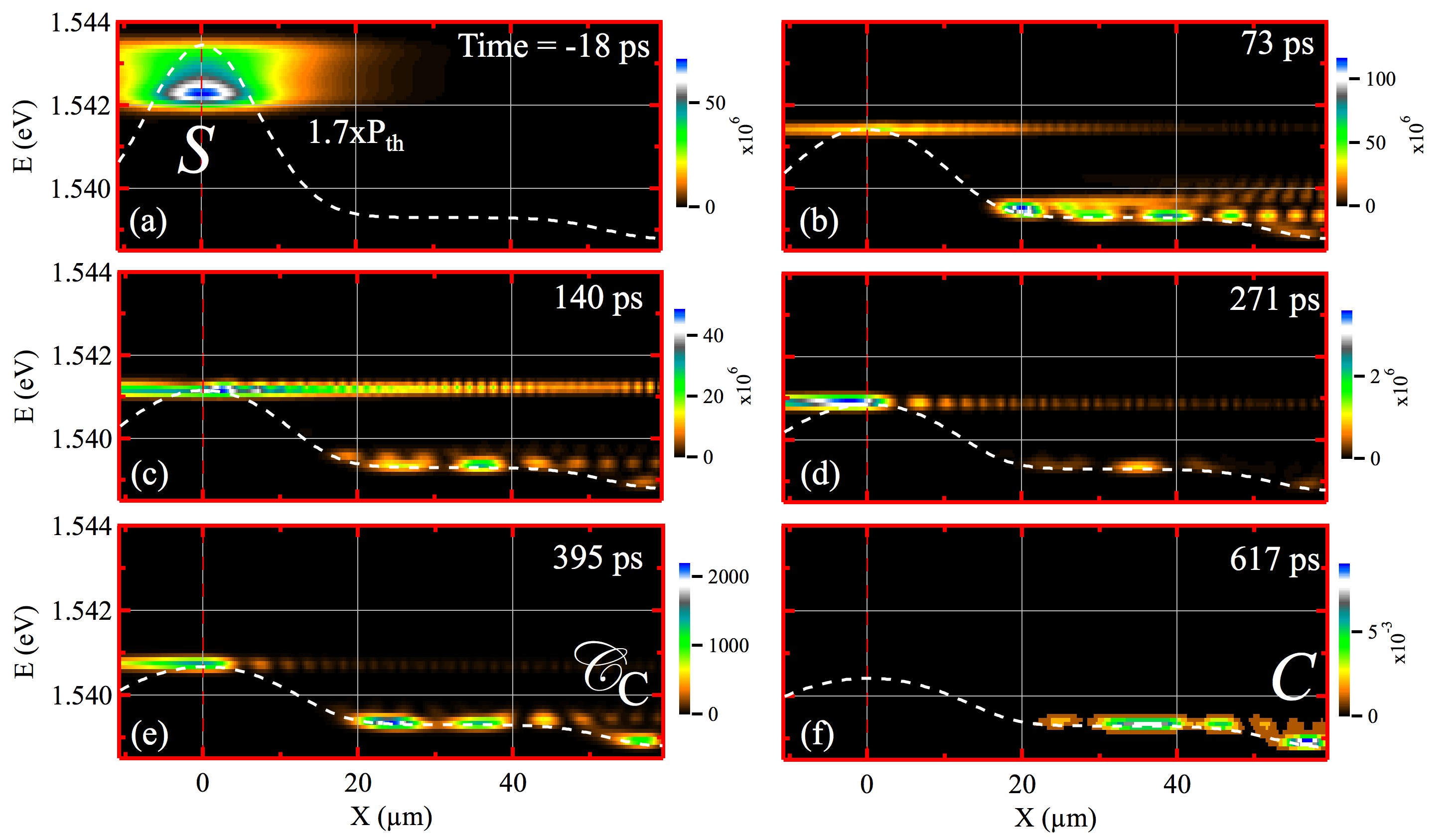}
\end{center}
\caption{(Color online) Energy vs. real space ($X$) for $P_S=1.7 \times P_{th}$ at different times shown by the labels. \emph{S}, \emph{C} and $\mathscr{C}_{C}$ mark the source, the collector and the trapped condensate at \emph{C} positions, respectively. The intensity is coded in a false color scale shown on the right of each panel.}
\label{fig:Time_2}
\end{figure}

For higher power ($P_S=1.7\times P_{th}$), Fig.~\ref{fig:Time_2} shows the onset of stimulated energy relaxation processes. As in the experimental case (Fig.~\ref{fig:fig3}) the relaxation takes place in two subsequent stages: first there is relaxation from the source state into the extended state with energy set by the ridge potential, followed by relaxation into the $\mathscr{C}_{C}$ condensate at low energy.
\begin{figure}
\setlength{\abovecaptionskip}{-5pt}
\setlength{\belowcaptionskip}{-2pt}
\begin{center}
\includegraphics[trim=0.3cm 0.2cm 0.5cm 0.3cm, clip=true,width=1.0\linewidth,angle=0]{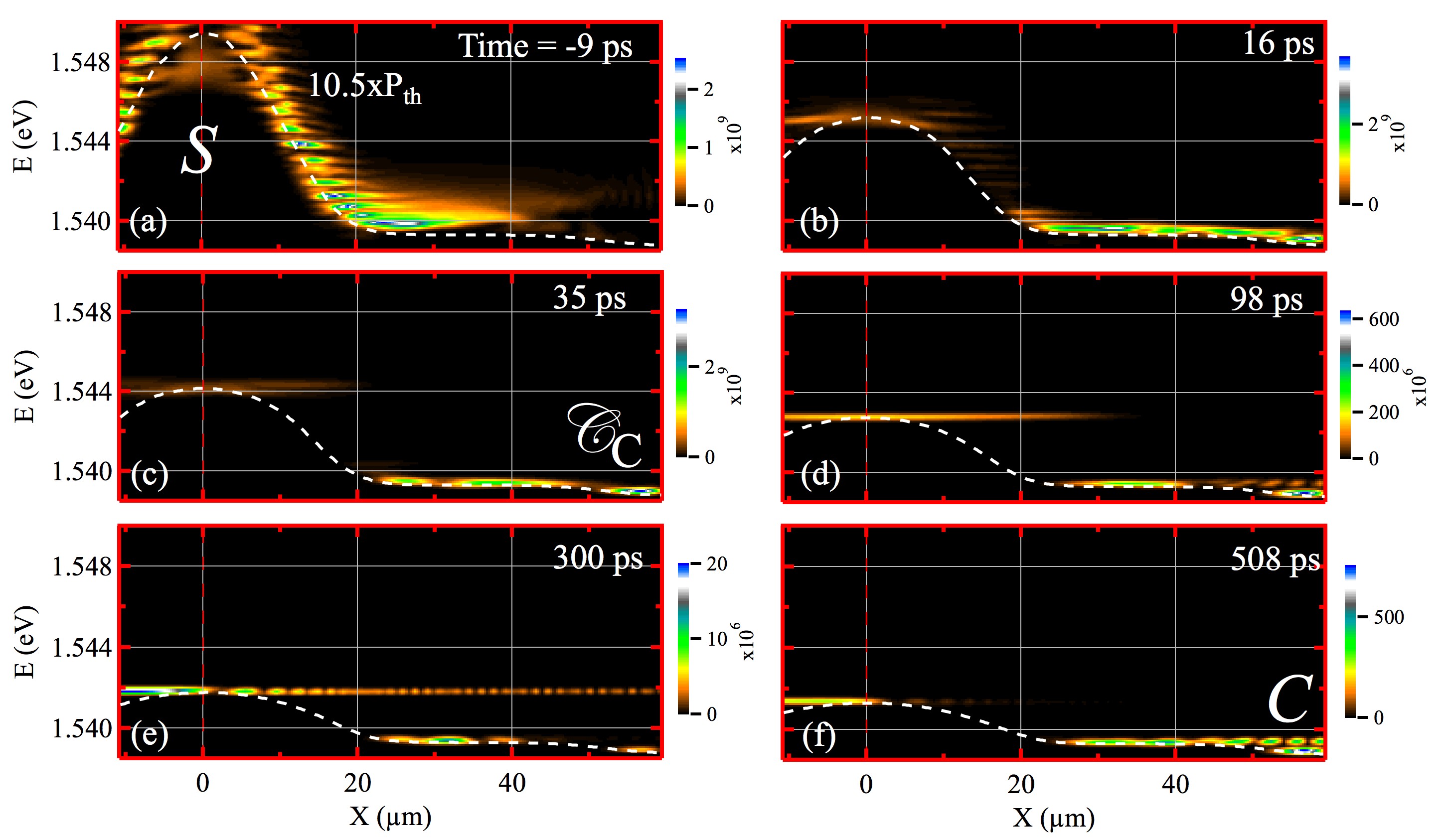}
\end{center}
\caption{(Color online) Energy vs. real space ($X$) for $P_S=10.5 \times P_{th}$ at different times shown by the labels. \emph{S}, \emph{C} and $\mathscr{C}_{C}$ mark the source, the collector and the trapped condensate at \emph{C} positions, respectively. The intensity is coded in a false color scale shown on the right of each panel.}
\label{fig:Time_3}
\end{figure}
At $10.5\times P_{th}$, Fig.~\ref{fig:Time_3} shows that the energy relaxation occurs rapidly. As in the experimental case (Fig.~\ref{fig:fig4}) the collector state is rapidly populated. It is interesting to note that, as shown for the experiments in Fig.~\ref{fig:fig12} (c), the blueshift of the condensate at the source position does not increase linearly with the pump power. This is because even though the injected hot exciton population can be expected to increase linearly, the increased carrier density results in a faster condensation rate due to the stimulation of scattering processes (hot exciton relaxation processes as well as processes that cause excitons to relax into condensed polaritons). Polaritons decay much faster than uncondensed hot excitons, such that a high intensity pumping of hot excitons is quickly depleted giving rise to a limited blueshift of polaritons at the source.

\subsection{Simulations for the two-beam excitation}
\label{subsec:sim_two_beam}

\begin{figure}
\setlength{\abovecaptionskip}{-5pt}
\setlength{\belowcaptionskip}{-2pt}
\begin{center}
\includegraphics[trim=0.3cm 0.2cm 0.5cm 0.4cm, clip=true,width=1.0\linewidth,angle=0]{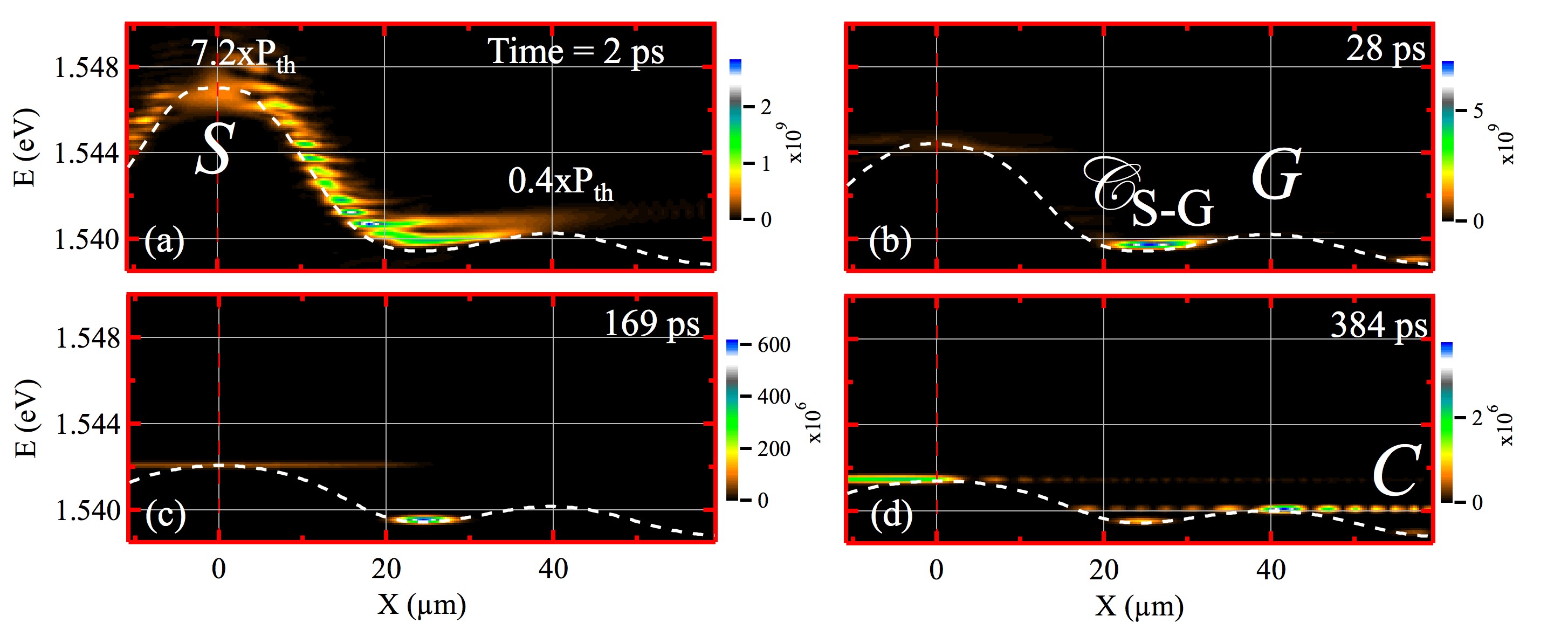}
\end{center}
\caption{(Color online) Energy vs. real space ($X$) for $P_S=7.2 \times P_{th}$ and $P_G=0.4\times P_{th}$ at different times shown by the labels. \emph{S}, \emph{G}, \emph{C} and $\mathscr{C}_{S-G}$ mark the source, the gate, the collector and the, between \emph{S-G}, trapped condensate positions, respectively. The intensity is coded in a false color scale shown on the right of each panel.}
\label{fig:Time_4}
\end{figure}

In the presence of the gating pulse, the propagation of the $\mathscr{C}_{S-G}$ condensate is blocked, as shown in Fig.~\ref{fig:Time_4}. This is due to the injected hot exciton density at the gate position that adds to the polariton effective potential profile, $V(x,t)$. At long times, the theory predicts a small transmission across the gate pulse, due to the decay of the potential barrier.

\subsection{Simulations on the power dependance of the energy/intensity decays}
\label{subsec:sim_power}

\begin{figure}
\setlength{\abovecaptionskip}{-5pt}
\setlength{\belowcaptionskip}{-2pt}
\begin{center}
\includegraphics[trim=0.3cm 0.2cm 0.4cm 0.3cm, clip=true,width=1.0\linewidth,angle=0]{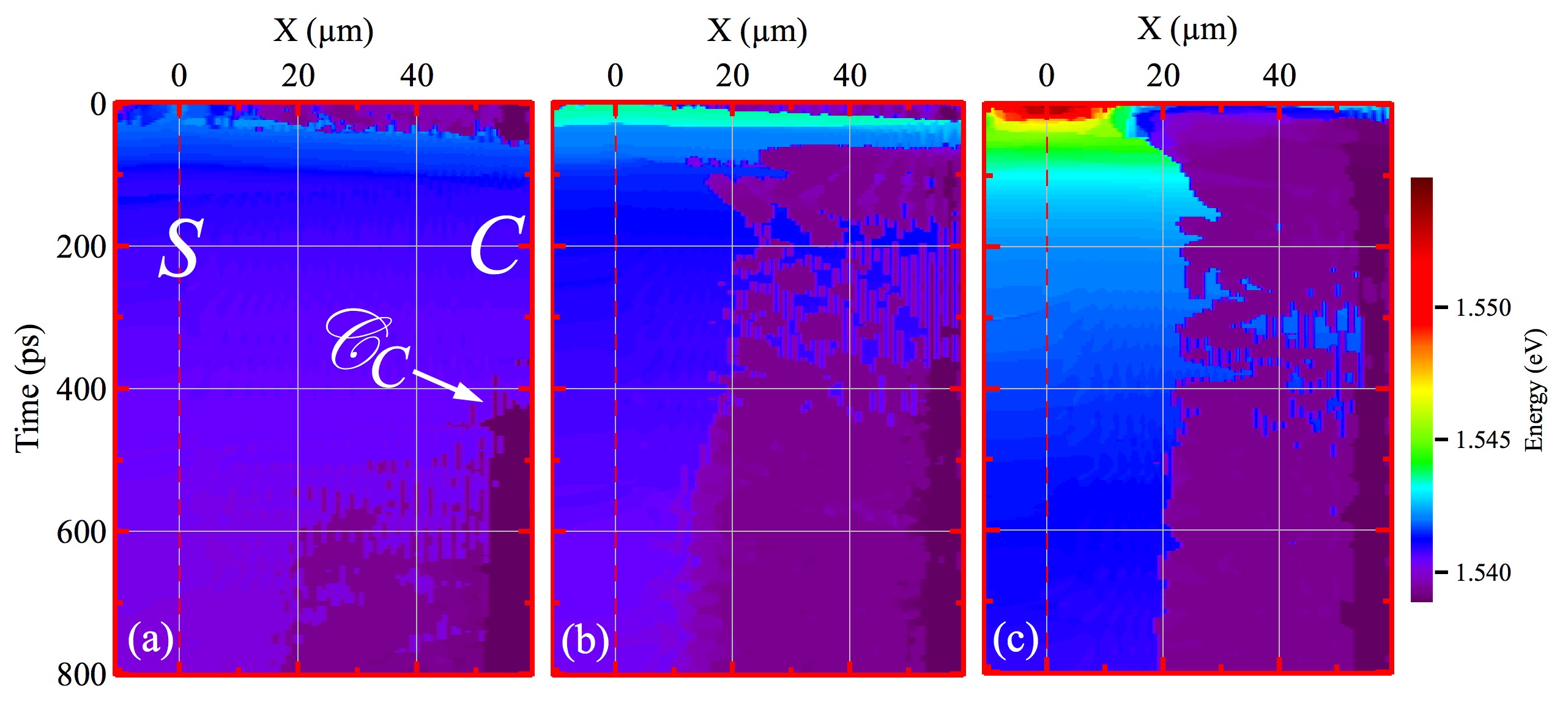}
\end{center}
\caption{(Color online) Energy of the emission vs. real space ($X$) and time. (a)$P_S=P_{th}$, (b) $P_S=1.7 \times P_{th}$, (c) $P_S=10.5 \times P_{th}$ and (d) $P_S=7.2 \times P_{th}$ and $P_G=0.4\times P_{th}$. \emph{S}, \emph{C} and $\mathscr{C}_{C}$ mark the source, the collector and the trapped condensate at \emph{C} positions, respectively. The intensity is coded in a false color scale shown on the right.}
\label{fig:Peak}
\end{figure}

Spatial-temporal maps of the peak emission energy with one-beam excitation are shown in Fig.~\ref{fig:Peak}. In panel (a) there is a fast propagation of a high energy mode from the source followed by a decrease in energy of the emission over the whole space. At longer times, one identifies relaxation into $\mathscr{C}_{C}$, near the wire edge. This behaviour is in qualitative agreement with the experimental result, however, it can be noted that the speed of propagation appears overestimated in the theory. This is because the theory neglects changes in the shape of the polariton dispersion caused by the hot-exciton induced blueshift, which can be particularly important at early times when the particles are strongly excitonic with a larger effective mass and slower group velocity. Figures~\ref{fig:Peak} (b) and (c) show the peak emission energy maps for increasing source intensity, where the relaxation into an extended state with lower energy than the source can be identified. The relaxation is stronger at the highest pump power, due to increased stimulated energy relaxation processes. This is also evidenced by the shorter time required for $\mathscr{C}_{C}$ to appear with increasing pump power.

\begin{figure}
\setlength{\abovecaptionskip}{-5pt}
\setlength{\belowcaptionskip}{-2pt}
\begin{center}
\includegraphics[trim=0.3cm 0.2cm 0.4cm 0.3cm, clip=true,width=1.0\linewidth,angle=0]{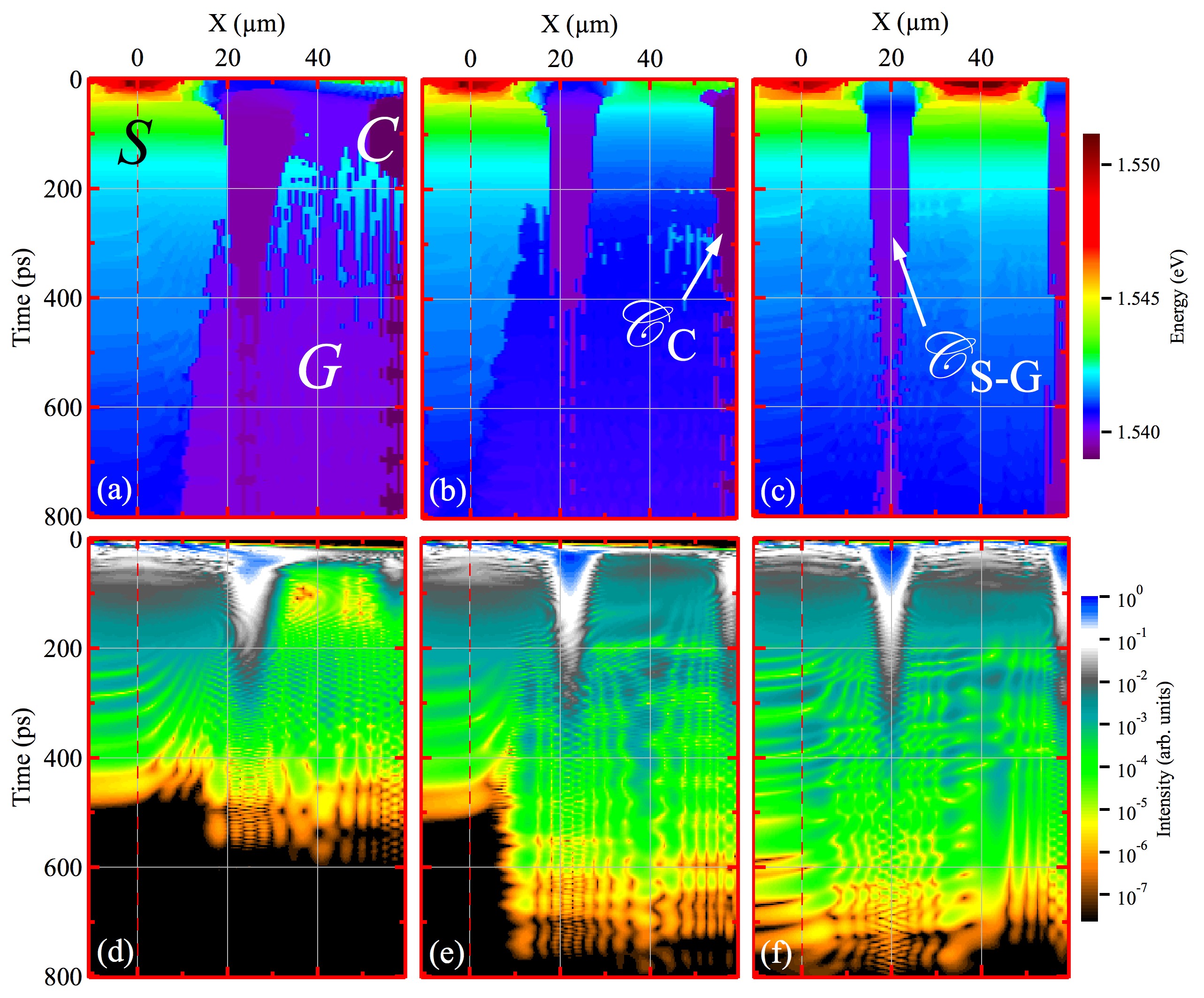}
\end{center}
\caption{(Color online) (a-c) Energy/(d-f) Intensity of the emission vs. real space ($X$) and time for a constant source excitation power $P_S=7.2\times P_{th}$ and different gate powers $P_G$: (a/d) 0.4$\times P_{th}$ (b/e) 1.8$\times P_{th}$ and (c/f) 9.0$\times P_{th}$. \emph{S}, \emph{G}, \emph{C}, $\mathscr{C}_{S-G}$ and $\mathscr{C}_{C}$ mark the source, the gate, the collector, the trapped condensate between \emph{S-G} and the trapped condensate at \emph{C} positions, respectively, see text for further details. The information is coded in a linear/logarithmic false color scale shown on the right side of the upper/lower row.}
\label{fig:Peak2}
\end{figure}

We discuss now the simulated energy- and intensity maps under two-beam excitation conditions compiled in Fig.~\ref{fig:Peak2}. Panel (a) shows the case when a  gate pulse, $P_G=0.4\times P_{th}$, is present. At short times the energy of the collector state is low, although it should be noted that it is populated with a negligible density, panel (d). The weak tunneling of particles across the gate is better evidenced by the increase of the collector state energy, since the tunneling particles have higher energy than the collector ground state. At these low gate powers, the theory appears to predict a high number of polaritons passing the gate. These polaritons have relatively high momentum and are expected to be less visible experimentally due to reduced photonic fractions. An increase of the gate power above threshold, panels (b,c,e,f), leads to an enhanced collector signal, as in the experimental case, see Figs.~\ref{fig:fig9} (b,c,e,f), and the device leaves the OFF state. This is expected as additional excited polaritons move directly from the gate to the collector. 

\begin{figure}
\setlength{\abovecaptionskip}{-5pt}
\setlength{\belowcaptionskip}{-2pt}
\begin{center}
\includegraphics[trim=0.3cm 0.2cm 0.5cm 0.4cm, clip=true,width=1.0\linewidth,angle=0]{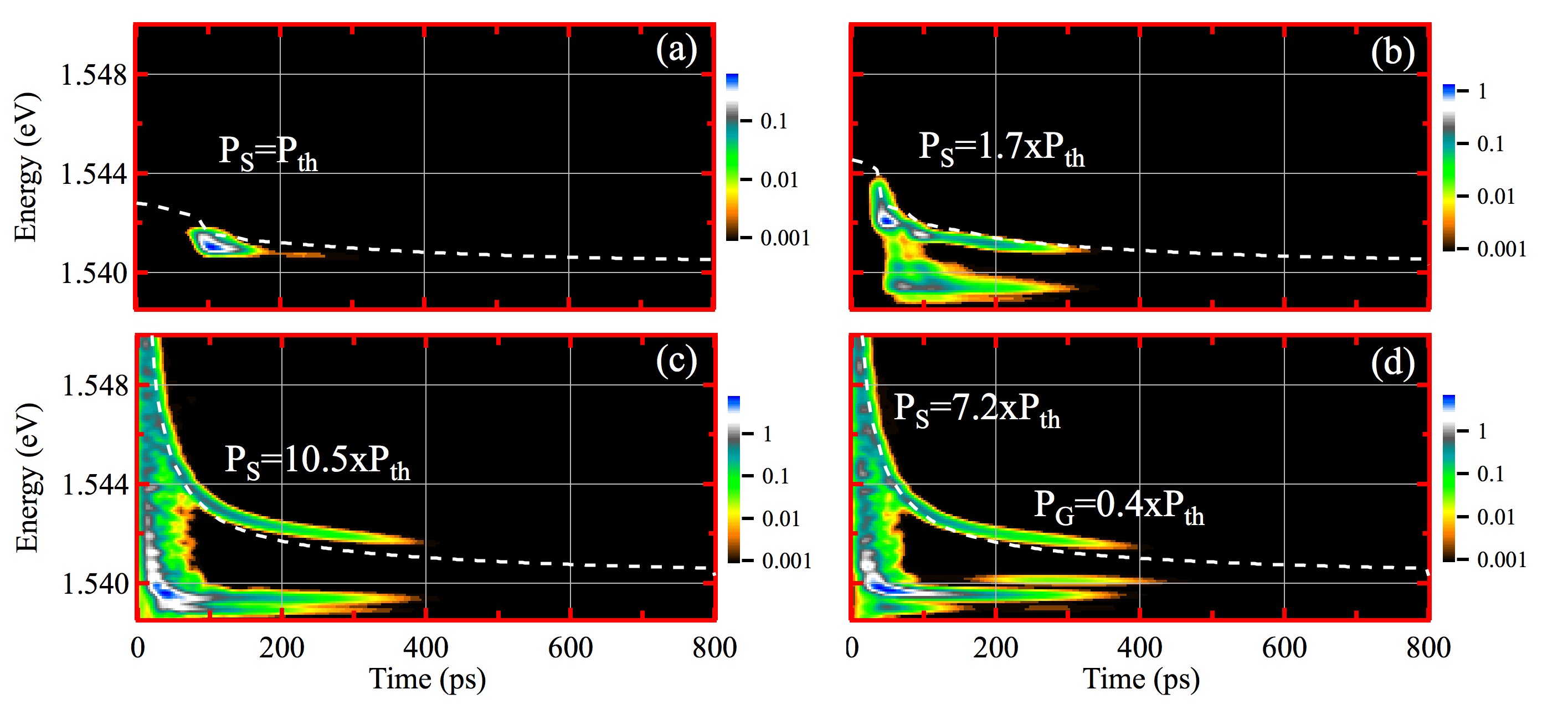}
\end{center}
\caption{(Color online) Energy decay, spatially-integrated, vs. time. (a)$P_S=P_{th}$, (b) $P_S=1.7 \times P_{th}$, (c) $P_S=10.5 \times P_{th}$ and (d) $P_S=7.2 \times P_{th}$ and $P_G=0.4\times P_{th}$. The intensity is coded in a false color scale shown on the right. The dashed curve shows the evolution of the polariton effective potential at the source, $V(x=0,t)$.}
\label{fig:It_all}
\end{figure}

Figure~\ref{fig:It_all} shows the time evolution of the spatially integrated spectra. In agreement with the experimental results (Fig.~\ref{fig:fig11}) there is a two timescale decay of the emission energy. At early times, the fast drop is due to the fast condensation rate in the presence of strong stimulated scattering. As mentioned earlier, this fast condensation rapidly depletes the hot exciton reservoir and the total particle density quickly drops as polaritons quickly decay. At longer times, reduced relaxation between the active and inactive reservoirs limits the effective condensation rate. The condensate is continuously fed while the reservoir intensities slowly decay.
The short lifetime of the condensate for pumping at threshold (Fig.~\ref{fig:It_all} (a)) is expected from the theoretical definition of threshold where the incoming rate $rN_A(x,t)$ is slightly larger than the polariton decay rate $\Gamma$ in Eq.~(\ref{eq:GP}). As soon as condensation starts, the reservoir density $N_A$ drops below threshold such that continued condensation cannot take place. For higher pump powers, Figs.~\ref{fig:It_all} (b,c) show that both emission into a high energy mode, corresponding to the source, and a lower energy emission coexist. This fact is in close agreement with the experimental data shown in Figs.~\ref{fig:fig11} (b,c) and (i), with the relaxation into the lower energy state occurring earlier for increased pumping power.

\section{Conclusions}
\label{sec:conclusions}

In summary, we have time-resolved the energy and intensity relaxation processes of excitons and polaritons in a microcavity ridge. Two different excitation configurations have been studied with one and two non-resonant, pulsed laser beams, permitting polariton condensate trapping on demand. A detailed analysis of the decay processes has been accomplished by mapping the energy and intensity emission along the ridge. Decay times of the source emission are reported under one-beam excitation, where we show the acceleration of the decaying processes as a function of increasing $P_S$. The time response of the polariton transistor switch is characterized and optimized.We used a generalized Gross-Pitaevskii model to describe the spatial dynamics of our propagating polariton condensates, which includes a phenomenological treatment of energy-relaxation processes that cause condensates to further thermalize as they travel in a non-uniform effective potential. The nonlinearity of energy-relaxation processes throughout the system, those causing relaxation between polariton states as well as relaxation between higher energy exciton states, is necessary to explain features of the experimental results. Approximating the system as a 1D system, we are able to describe the main qualitative features of the experiment. While the system is essentially 1D, lateral expansion could result in a lower propagation speed than that predicted theoretically. We intend to investigate lateral propagation effects in future work.

The optimization of individual condensate transistor elements, as we have reported here, is an essential step towards developing information processing devices with the present scheme. In the future, an important goal is the achievement of cascadability and fan-out of multiple elements for the construction of extended circuits. Such a feat was very recently achieved in polariton based systems with coherent near-resonant excitation.~\cite{Ballarini2012} Achieving the same with the gating of incoherently generated polariton condensates, as we study here, would be particularly promising as it would open up routes toward electrically injected devices and consequently hybrid electro-optical processing systems.

\acknowledgments

C.A. and G.T. acknowledge financial support from Spanish FPU and FPI scholarships, respectively. P.S. acknowledges Greek GSRT program ``ARISTEIA" (1978) for financial support. The work was partially supported by the Spanish MEC MAT2011-22997, CAM (S-2009/ESP-1503) and FP7 ITN's ``Clermont4" (235114), ``Spin-optronics" (237252) and INDEX (289968) projects.

\section*{Appendix A: List of symbols}
\label{sec:symbols}

In this Appendix we define the symbols used as abbreviations along the manuscript.

\begin{table}[h]
\centering
\begin{tabular*}{1\columnwidth}{@{\extracolsep{\fill}}|c|>{\centering}p{0.8\columnwidth}|}
\hline
\bf{Symbol} & \bf{Meaning}\tabularnewline
\hline
\hline 
$\Delta E_S$ & Energy shift of the emission at the Source position\tabularnewline
\hline
$\tau_{fast}$ & Energy fast decay time at the Source\tabularnewline
\hline
$\tau_{slow}$ & Energy slow decay time at the Source\tabularnewline
\hline
\emph{C} & Collector\tabularnewline
\hline 
$\mathscr{C}_C$ & Polariton condensate trapped at the Collector\tabularnewline
\hline
$\mathscr{C}_{S-G}$ & Polariton condensate trapped between Source and Gate positions\tabularnewline
\hline
$\mathscr{D}^P$ & Horizon of propagating polaritons along the ridge\tabularnewline
\hline
$\mathscr{D}^{S-P}$ & Horizon given by the interface between carries at the Source and propagating polaritons\tabularnewline
\hline
\emph{G} & Gate\tabularnewline
\hline 
$I\left(\mathscr{C}_{C}\right)$ & Emission intensity of the condensate trapped at the Collector\tabularnewline
\hline
$I\left(\mathscr{C}_{S-G}\right)$ & Emission intensity of the condensate trapped between Source and Gate\tabularnewline
\hline
$I\left(S\right)$ & Emission intensity at the Source position\tabularnewline
\hline
$P_G$ & Gate beam power\tabularnewline
\hline
$P_S$ & Source beam power\tabularnewline
\hline
$P_{th}$ & Pump power threshold for polariton condensation\tabularnewline
\hline
\emph{S} & Source\tabularnewline
\hline 
$t_{down}$ & Time time for the Collector intensity to drop from 1 to 0.5\tabularnewline
\hline
$T_{ON}$ & Time delay between maxima of the Source and Collector intensities\tabularnewline
\hline
$t_{up}$ & Time time for the Collector intensity to rise from 0.5 to 1\tabularnewline
\hline
$v^P$ & Mean speed of propagating polaritons \tabularnewline
\hline
$v^S$ & Mean speed of carriers at the Source position\tabularnewline
\hline

\end{tabular*}

\caption{List of symbols used to describe the experiments.}
\label{tablexp}
\end{table}


\begin{table}[h]
\centering
\begin{tabular*}{1\columnwidth}{@{\extracolsep{\fill}}|c|>{\centering}p{0.8\columnwidth}|}
\hline
\bf{Symbol} & \bf{Meaning}\tabularnewline
\hline
\hline
$\alpha$ & Polariton-polariton interaction strength\tabularnewline
\hline
$\Gamma$ & Polariton decay rate\tabularnewline
\hline
$\Gamma_A$ & Decay rate of the active exciton reservoir\tabularnewline
\hline
$\Gamma_D$ & Decay rate of the dark exciton reservoir\tabularnewline
\hline
$\Gamma_I$ & Decay rate of the inactive exciton reservoir\tabularnewline
\hline
$\nu$/$\nu^\prime$ & Phenomenological parameters for the strength of energy relaxation\tabularnewline
\hline
$\psi(x,t)$ & Polariton wavefunction\tabularnewline
\hline
$E_{LP}$ & Lower polariton branch energy dispersion\tabularnewline
\hline
$g_D$ & Polariton-exciton interaction strength for dark excitons \tabularnewline
\hline
$g_I$ & Polariton-exciton interaction strength for indirect excitons \tabularnewline
\hline
$g_R$ & Polariton-exciton interaction strength for reservoir excitons \tabularnewline
\hline
$m$ & Polariton effective mass\tabularnewline
\hline
$m_e$ & Free electron mass\tabularnewline
\hline
$N_A$ & Density distribution of active excitons\tabularnewline
\hline
$N_I$ & Density distribution of inactive excitons\tabularnewline
\hline
$N_D$ & Density distribution of dark excitons\tabularnewline
\hline

$r$ & Polariton condensation rate\tabularnewline
\hline
$t_D$ & Linear coupling to the dark exciton reservoir\tabularnewline
\hline
$t_R$ & Linear coupling to the inactive exciton reservoir\tabularnewline
\hline
$t_R^\prime$ & Non-linear coupling to the inactive exciton reservoir\tabularnewline
\hline
$V\left(x,t\right)$ & Effective polariton potential\tabularnewline
\hline
$V_0\left(x\right)$ & Wire structural potential\tabularnewline
\hline
\end{tabular*}

\caption{List of symbols used on Section~\ref{sec:model} Model.}

\label{tableteo}

\end{table}

%

\begin{thebibliography}{10}
\bibitem{energynote}
The energy maps consider a cut-off based on the experimental background noise,
  leaving white points when the intensity is below that noise level.
\end{thebibliography}

\end{document}